\documentclass[11pt]{article}
\usepackage[pdftex,colorlinks=true]{hyperref} 
\usepackage{epsfig}
\usepackage{amsmath}    
\usepackage{amssymb}
\usepackage{graphicx}
\usepackage{amsfonts}   
\usepackage{latexsym}   
\usepackage{slashed}
\usepackage{yfonts} 
\usepackage{verbatim}
\usepackage{wrapfig}
\usepackage{mathrsfs}
\usepackage{cancel}

\global\newcount\itemno \global\itemno=0

\def\itemaut#1{\global\advance\itemno by1\noindent\item{\the\itemno.}#1}

\newif{\ifeq}		
\eqfalse		
\newcommand{\be}{\begin{equation}}
\newcommand{\ee}{\end{equation}}
\newcommand{\bes}{\begin{equation*}}
\newcommand{\ees}{\end{equation*}}
\newcommand{\bea}{\begin{eqnarray}}
\newcommand{\eea}{\end{eqnarray}}
\newcommand{\beas}{\begin{eqnarray*}}
\newcommand{\eeas}{\end{eqnarray*}}

\def\({\left(}
\def\){\right)}
\def\[{\left[}
\def\]{\right]}

\def\frac#1#2{{#1 \over #2}}

\def\vev#1{\langle{#1}\rangle}

\newcommand{\half}{\frac{1}{2}}

\renewcommand{\b}{\beta}

\renewcommand{\r}{\rho}
\newcommand{\s}{\sigma}

\newcommand{\w}{\omega}
\newcommand{\m}{\mu}

\newcommand{\eps}{\epsilon}

\newcommand{\CD}{{\cal D}}

\newcommand{\CF}{{\cal F}}

\newcommand{\CJ}{{\cal J}}

\newcommand{\CL}{{\cal L}}

\newcommand{\CO}{{\cal O}}

\def\eg{{\it e.g.}}
\def\ie{{\it i.e.}}

\newcommand{\lsim}{\,\raise.3ex\hbox{$<$\kern-.75em\lower1ex\hbox{$\sim$}}\,}
\newcommand{\gsim}{\,\raise.3ex\hbox{$>$\kern-.75em\lower1ex\hbox{$\sim$}}\,}

\def\p{\partial}

\def\II{\relax{I\kern-.10em I}}

\usepackage{amssymb}

\numberwithin{equation}{section}

\begin{document}

\begin{titlepage}
\begin{flushright}MIT-CTP-4230 \end{flushright}
\vskip 1in

\begin{center}
{\Large{Towards a Non-Relativistic Holographic Superfluid}}
\vskip 0.5in Allan Adams and Juven Wang
\vskip 0.4in {\it Center for Theoretical Physics\\  Massachusetts Institute of Technology\\ Cambridge, MA  02139, USA}
\end{center}
\vskip 0.5in

\begin{abstract}\noindent
We explore the phase structure of a holographic toy model of superfluid states in non-relativistic conformal field theories.
At low background mass density, we find a familiar 
second-order 
transition 
to a superfluid phase 
at finite temperature.  
Increasing the chemical potential for the probe charge density drives this transition strongly first order as the low-temperature superfluid phase merges with a thermodynamically disfavored high-temperature condensed phase.
At high background mass density, the system reenters the normal phase as the temperature is lowered further, hinting at a zero-temperature quantum phase transition as the background density is varied.
Given the unusual thermodynamics of the background black hole, however, it seems likely that the true ground state is 
another configuration altogether.
\end{abstract}

\end{titlepage}

\section{Introduction}\label{Sec:Intro}

Non-relativistic superfluids provide a high-precision laboratory in which to probe many-body physics in the extreme quantum regime \cite{zwerger}.  In an effort to bring the tools of holography \cite{Maldacena:1997re,Gubser:1998bc,Witten:1998qj} to bear on these systems, considerable effort has been devoted to studying non-relativistic deformations of relativistic examples\footnote{Since the non-relativistic conformal group is a subgroup of the relativistic group in one higher dimension, we can construct a non-relativistic conformal field theory (NRCFT) by turning on an operator in a relativistic conformal field theory (CFT) which breaks the relativistic group to its non-relativistic subgroup.  Taking the operator to be marginal in the NRCFT \cite{Mehen:1999nd,Nishida:2007pj} requires it to be irrelevant in the CFT.  Holographically, this corresponds to a 1-parameter deformation of the geometry which alters the asymptotic geometry from Anti de Sitter (AdS), whose isometries form the relativistic conformal group, to Schr\"odinger \cite{Son:2008ye, Balasubramanian:2008dm}, whose isometries fill out the non-relativistic conformal group.} which enjoy $z=2$ scaling \cite{Son:2008ye, Balasubramanian:2008dm,Adams:2008wt,Herzog:2008wg,Maldacena:2008wh}.  Unfortunately, such deformations generate highly atypical states in the resulting NRCFT whose thermodynamic and other properties are tightly constrained by their relativistic births.  In particular, they are in general far from the superfluid groundstates of the corresponding systems, for which we currently  have no description.

In this paper we examine certain superfluid states in a holographic NRCFT in a probe approximation.  
Our strategy is essentially the same as in the AdS case (see \eg\ \cite{Gubser:2008px, Hartnoll:2008vx}): we study an Abelian-Higgs theory in the background of a neutral asymptotically-Schr\"odinger black hole \cite{Adams:2008wt, Herzog:2008wg, Maldacena:2008wh}\ in the probe approximation.  Several features of the geometry, however, make the resulting analysis qualitatively different.  For example, we are now forced to turn on two components of the bulk gauge field: $A_{t}$, dual to a boundary charge current, and $A_{\xi}$, dual to a boundary Mass\footnote{%
In an NRCFT, each primary operator is characterized by not only a Dimension $\Delta$ but also by its Mass, $M$, where
Mass is the name of a central extension $\hat{M}$ in the NR conformal group.  In the case of free fermions, $\hat{M}=M\psi^{\dagger}\psi$, hence it is often called the ``Number'' operator -- we prefer ``Mass'' to disambiguate the various meanings of ``number''.} current.  By itself, this is not a big deal.  What's surprising given intuition from the relativistic case is that the boundary value of this second vector component, $M_{o}=A_{\xi}|_{\p}$, weasels its way into the {\em dimension} of the boundary order parameter as $\Delta = 2\pm \sqrt{4+m^2+q^{2}M_{o}^{2}}$.  Specifying the boundary NRCFT thus requires not just specifying the bulk matter fields and their interactions, but also the asymptotic fall-offs of some of the bulk fields.  Similar effects arise in the holographic renormalization of the theory, which as usual requires introducing counterterms which depend on the boundary operator dimensions; here, these counterterms will explicitly depend on the boundary values of some bulk fields, too (see \eg\ \cite{Guica:2010sw}\ for a discussion of such effects).

To build a truly NR superfluid, then, we must generate a condensate for a boundary operator with non-zero Mass eigenvalue, $M\neq0$.  This is the role of the second component of the gauge field -- in a gauge where the phase of the condensate is constant, the Mass eigenvalue is simply $M=-q\,M_{o}$.   The boundary value of the second component of the gauge field thus controls the breaking of the Mass symmetry in the superfluid phase.

While the background about which we perturb is a 1-parameter deformation of a relativistic example, the superfluid state we find is not, and indeed enjoys quite distinct phenomenology from its AdS cousins.  Fundamentally, the non-relativistic  condensate is characterized by one more quantum number than in the relativistic case -- the Mass eigenvalue, $M$, of the order parameter -- with the NR condensate breaking the symmetry generated by the Mass operator, a key signature of a non-relativistic superfluid.  As we shall see, this leads to a host of interesting effects in the strongly NR regime, including the appearance of a thermodynamically unstable high-temperature condensed phase which drives the superconducting transition from 2$^{nd}$ order to 1$^{st}$ at a multicritical point, the persistence of a condensate even in the absence of a chemical potential for the charge density,  and reentrance of the normal phase at low temperatures for sufficiently large background density.

It is tempting to interpret this re-entrance as signaling a zero temperature quantum phase transition as the background mass density is tuned.  However, the re-entrant normal state is again the simple 1-parameter deformation which we do not expect to be the true equilibrium groundstate, so we do not expect this probe analysis to be the end of the story.  Meanwhile, it remains possible that the system is in fact reentrant for all values of the background Mass density as $T\to0$, where our probe approximation becomes unreliable.  Resolving these puzzles, however, requires going beyond the truncated probe approximation discussed in this paper; we leave them to future study.

The plan of the paper is as follows.
In Section 2 we quickly describe the basic strategy and computational setup, with various details elaborated in Appendices.
In Section 3 we explore the phenomenology and phase stucture of holographic superfluids outside a Schr\"odinger black hole (an analogous study in the background of a Schr\"odinger soliton \cite{Mann:2009xk}\ is performed in Appendix A -- while this is not in the same ensemble as the black hole, it provides an alternate example with surprising physics of its own).
We close in Section 5 with a summary and list of next steps.

\section{The Setup}\label{Sec:Probe}

Our basic strategy involves studying an Abelian Higgs system,
\be\label{EQ:ProbeAction}
\CL_{probe} ={1\over e^{2}}\( -{1\over4} F^{2} -|\CD\Phi|^{2}  -m^{2}|\Phi|^{2}  \) \,,
\ee
as a perturbation around the planar Schr\"odinger black hole background,
\be\label{EQ:pappalardoBH}
ds^2 = 
\(-f + {(f-1)^{2}\over 4(K-1)}\){dt^{2}\over K r^{4}}
+ 
{1+f \over r^{2} K} dt \, d\xi 
+{K-1\over K}d\xi^{2}
+ {d\vec{x}^{2} \over r^{2}}
+ {dr^{2}  \over f \, r^{2}}\,.
\ee
in the probe limit, $e^{2}\to\infty$.  Here, 
$f=1-r^{4}(\pi T\Omega)^{4/3}$,  
$K=1+r^2 \Omega^2$
and the metric is given in string frame.  One can think of this as a rather extreme truncation of the charged Schr\"odinger black hole system \cite{Adams:2009dm,Imeroni:2009cs}\ where we drop the coupling of the vector to the scalar and massive vector of the black hole background, or simply as a holographic toy model.  The geometry is controlled by two physical parameters, the background mass density, $\Omega$, and the temperature, $T$,
with the horizon located at the radial coordinate $r_{H}=(\pi T\Omega)^{-1/3}$.  

For spatially homogeneous solutions, we can without loss of generality set $\vec{A}=0$ and take 
$\Phi = \phi(r)$ 
and $A = A_{t}(r)dt + A_{\xi}(r) d\xi$.  
In Einstein frame, the equations of motion take the form,
\bea\label{EQ:EOM1}
&& \hspace{-.65cm}
f^{2}r^{2}\phi'' - f(4-f)r\phi' 
-\[\vphantom{\frac{A^{2}}{B_{H}}^{2}}
f\(q^{2}A_{\xi}^{2}
+2q^{2} r^{2}A_{\xi}A_{t}
+m^{2}K^{1/3}\)\right. \nonumber\\
&& \hspace{4cm}\left. -{q^{2}(f-1)^{2}\over 4(K-1)}\(A_{\xi} - 2r_{H}^{4}\Omega^{2}A_{t}\)^{2}
\vphantom{\frac{A^{2}}{B_{H}}^{2}}\]\phi =0~~~\\
\label{EQ:EOM2}
&& \hspace{-.65cm}f r^{2}A_{t}'' - \(2-{f\over3}(7K-4)\){r\over K}A_{t}' 
- \(2+f(f-1)+{(f-1)^{2}\over K-1}\){1\over Kr}A_{\xi}' -2q^{2}K^{1/3}  \phi^{2} A_{t} 
= 0~~~~~~~~~~~\\
\label{EQ:EOM3}
&& \hspace{-.65cm}f r^{2}A_{\xi}'' - \(4K-2-{2+K\over3}f\){r\over K}A_{\xi}' -4(K-1){r^{3}\over K}A_{t}'  -2q^{2} K^{1/3}\phi^{2} A_{\xi}  
= 0 ~~~
\eea
Note that $A_{\xi}\neq A_{t}$.

\subsection{Asymptotic Behavior and the Holographic Dictionary}

Near the boundary at $r=0$, the vector components behave as,
\be
A_t    = \mu_{Q} +\r_{Q}\, r^2+\dots\,, ~~~~~~~~~~
A_\xi = M_{o} + \r_{M}\, r^2 + \dots
\ee
where the various $\dots$ represent various (possibly non-normalizable) terms whose coefficients are entirely fixed by the equations of motion and the values of these integration constants, $\mu_{Q}$, $\r_{Q}$, $M_{o}$ and $\r_{M}$.\footnote{In particular, the leading term for $A_{t}$ runs as $-2\rho_{M}\log(r)$.  While formally the dominant term, it is determined by the equations of motion and $\r_{M}$ and thus does not represent an independent mode of the system.  Importantly, due to factors of the inverse metric, this log running does not lead any components of the bulk stress tensor to diverge.  A complete holographic renormalization of this system would settle the dictionary, but is beyond the scope of the present paper; for the moment we simply take the above dictionary as a provisional interpretation which is supported by the consistency of the results below.  Interestingly, while $A_{t}$ has no log in fully backreacted charged-black-hole solutions \cite{Adams:2009dm,Imeroni:2009cs}, linearizing the Maxwell equation around these solutions does generate a $\log$ without changing any other of the asymptotics of the vector, so this log is likely a simple consequence of an extreme truncation of the full charged black hole system.  It would be interesting to study the full system and see what, if anything, changes.}\
As usual, $\mu_{Q}$ represents the chemical potential per unit charge, which effectively sets the zero of energy in the boundary theory -- the gauge-invariant bulk quantity that becomes the boundary hamiltonian acting on the operator dual to the bulk matter field of charge $q$ is $(i\p_{t}+q\,A_{t})$; at the boundary, for plane waves $e^{-i\w t}$, this becomes $(\w+q\,\m_{Q})$.
Thus, one insertion of the charged operator $\CO \,e^{-i\w t}$ costs $\delta E = (\w+q\,\mu_{Q})$.
As usual, $\rho_{Q}$ computes the induced charge density.  

It might be tempting to think of $M_{o}$ as a chemical potential for the Mass operator, $\hat{M}$.  However, this is not quite right -- it is a superselection parameter.  Recall that, holographically, 
\be
\hat{M} \equiv \hat{P}_{\xi}|_{\p}= -i(\p_{\xi}-iqA_{\xi})|_{\p}\,,
\ee
\ie\ $\hat{M}$ is the boundary value of the gauge-invariant $\xi$-momentum in the bulk.
The mass eigenvalue of a boundary operator dual to a bulk field with $\xi$-momentum $\ell$ and charge $q$ is thus $M=(\ell-q\,M_{o})$, where $M_{o}=A_{\xi}|_{\p}$.
Like a chemical potential, $M_{o}$ sets a bias for the mass $M$, shifting it away from its $\xi$-momentum, $\ell$.  But the mass in an NRCFT is not a parameter, it is part of the definition of the theory.  Thus, once we fix gauge in the bulk, {different values of $A_{\xi}|_{\p}$ correspond to distinct NRCFTs}, not to a fixed theory with different background fields turned on.  In particular, as we will see momentarily, the dimensions of various boundary operators depend on $A_{\xi}|_{\p}$, an unfamiliar effect.  $\rho_{M}$ computes the Mass density coupled to $A_{\mu}$.  Henceforth we fix gauge in the bulk such that $\ell=0$ and $M = -qA_{\xi}|_{\p}$.


As for our charged scalar, near the boundary at $r\to0$ it behaves as
\be
\phi   \sim \phi_1r^{\Delta_-} +\phi_2r^{\Delta_+}+\dots\,,
\ee
where 
\be
\Delta_\pm
=2\pm \sqrt{4+m^2+q^{2}M_{o}^2} \, .
\ee
(Note that we will occasionally write $\Delta_{1}$ and $\Delta_{2}$ for $\Delta_{-}$ and $\Delta_{+}$, respectively.)  In the window $1\!<\!\Delta_{-}\!<\!2$, both components $\phi_{1,2}$ are normalizable, so we may interpret either of $\phi_{1,2}$ as the vev $\vev{\CO}$, with the other representing the source $\CJ$.  These two choices correspond to alternate quantizations of the boundary NRCFT \cite{Son:2008ye,Klebanov:1999tb}.  
We will focus on the choice $\vev{\CO}\propto\phi_{1}$ and $\CJ\propto\phi_{2}$ for reasons which will become clear in the next section.

Importantly, the dimensions, $\Delta_{\pm}$, depend not only on the mass of the bulk scalar, as in AdS,
but also on the boundary value of a bulk field, $M_{o}=A_{\xi}|_{\p}$.    As discussed above, by the holographic relation $\hat{M} \equiv -i(\p_{\xi}-iqA_{\xi})|_{\p}$, this quantity is nothing but the Mass eigenvalue of the dual operator, 
$$
M=-qA_{\xi}|_{\p} = -q\,M_{o}.
$$
Our expression for the dimensions above then becomes,
$
\Delta_{\pm} = 2\pm \sqrt{4+m^2+M^2}\,,
$
which is the expected form \cite{Son:2008ye,Balasubramanian:2008dm}, including the quadratic dependence on $M$ inside the radical.

Now, as discussed in  \cite{Adams:2008wt,Herzog:2008wg,Maldacena:2008wh}, the free energy of the full system takes the form, $F \equiv E+\m_{M}\hat{M}$, where $\m_{M}={-1\over2\Omega^{2}r_{H}^{4}}$ is determined by the background spacetime.  The total free energy per insertion of an operator dual to a bulk field with $\xi$-momentum $\ell$, frequency $\w$, and coupled with charge $q$ to our gauge field is thus $\delta F = (\w+q\,\mu_{Q})+\mu_{M} M$, where $M=(\ell-q\,M_{o})$

\subsection{Near-Horizon Behavior and Setting Up the Calculation}

In the bulk, we are thus left with a six-parameter family of solutions labeled by sources $\(\mu_{Q},\,M_{o},\,{\cal J}\)$ and responses $\(\r_{Q},\,\r_{M},\,\vev{\CO}\)$.   Holographically, we expect boundary conditions at the horizon, where the radial equations of motion degenerate, to impose three additional constraints. Together with the two parameters $T$ and $\Omega$ of the background geometry, this should leave us with a five-parameter phase space.  To verify this, we need to study the behaviour of our solutions near the horizon.

The equations of motion degenerate at the black hole horizon, so we must impose boundary conditions to pick the appropriate solutions.  As usual, it suffices to impose regularity at the horizon, which is in any case necessary for the validity of the probe approximation.  Assuming regularity, the equations of motion as presented in (\ref{EQ:EOM1}) -- (\ref{EQ:EOM3}) degenerate into three algebraic equations relating the six horizon values of the fields and their derivatives, as expected,
\bea
\(A_{\xi}(r_{H})- 2r_{H}^{4}\Omega^{2}A_{t}(r_{H})\)^{2} \phi(r_{H}) &=& 0 \\
- 2r_{H}^{2}A_{t}'(r_{H}) + \({2K_{H}-1\over K_{H}-1}\)A_{\xi}'(r_{H}) + 2q^{2}r_{H}K_{H}^{4/3}\phi^{2}(r_{H})A_{t}(r_{H}) &=& 0 \\
4r_{H}\phi'(r_{H}) +\( 4K_{H}(K_{H}-1)r_{H}^{4}q^{2}A_{t}^{2}(r_{H})   +{m^{2}K^{1/3}}\)\phi(r_{H})  &=&0\,.
\eea

This suggests a simple numerical strategy for constructing superfluid states of our holographic NRCFT.  To specify a solution to the full equations of motion, we fix any three of $\mu_{Q}$, $M_{o}$, $\CJ$, $\r_{Q}$, $\r_{M}$ and $\vev{\CO}$ at the boundary and impose the above regularity conditions at the horizon. Since we are interested in spontaneously generated condensates, we will generally set $\CJ=0$.  The resulting two-point boundary value problem can be solved numerically in various ways.  The most straightforward is a brute-force shooting method, as typically employed in the relativistic case.

In sweeping out parameter space, however, we must be careful to vary the {\em parameters} of the NRCFT while holding the NRCFT itself fixed -- \ie, while holding the spectrum of quantum numbers fixed.  This is straightforward in AdS, where fixing the set of dimensions reduces to fixing the bulk mass $m^{2}$ of the bulk scalar.  Here, however, the dimension $\Delta$ and Mass $M$ of the boundary scalar operator depend on the asymptotic value of $A_{\xi}$ as $M = -q\,A_{\xi}|_{\p}$ and $\Delta_{\pm} = 2\pm \sqrt{4+m^2+M^2}$.  Before sweeping out parameter space, then, we must fix $A_{\xi}|_{\p}$=$M_{o}$.  
As we've already set $\CJ=0$, fixing the system thus leaves us with a three-parameter phase space labeled by $\mu_{Q}$, $\Omega$ and $T$.

This peculiar behavior -- that the definition of the boundary CFT depends on the boundary behavior of the bulk fields -- is a very general phenomenon in Schr\"odinger holography.  Indeed, renormalizing the boundary stress tensor, say, or other operators in the boundary NRCFT, requires counterterms which are local in time and space, but which depend explicitly on dimensions, $\Delta$, and thus on the asymptotic values of $A_{\xi}$, in a mildly non-local fashion.  Such field-dependent counterterms have appeared previously in attempts to renormalize holographic NRCFTs, most recently in  \cite{Guica:2010sw}.  A complete understanding of the holographic renormalization of these theories is clearly of considerable interest.

\subsection{Conductivity}

We can compute the conductivity in our superconducting background by studying linear response to a time-dependent vector potential $A_{x}\sim e^{-i\w t}$.  As usual, this boils down to solving the equation of motion for the bulk gauge component $A_{x}$ linearized about the superfluid background and subject to infalling boundary conditions at the horizon.  Setting $A_{x}=a(x)e^{-i\w t}$, we have
\be\label{EQ:Cond}
f r^{2}a_{x}'' - \(4-f{2+7K\over3K}\)r a_{x}' +\({\w^{2}r^{4}(K-1)\over f} - 2q^{2} K^{1/3}\phi^{2}\)a_{x} = 0\,.
\ee

Near the horizon, this reduces to,
\be
(\omega^2+(\frac{4 \epsilon}{r_H^3 \Omega} \frac{d}{d \epsilon})^2 ) a_x(\epsilon) \simeq 0
\ee
where $r=r_{H}-\eps$.  The infalling solutions thus takes the form,
\be
A_{x} = a_{0}\,e^{-i\w t}(r-r_{H})^{-i\w\over4\pi T}(1+a_{1}(r-r_{H})+\dots)
\ee

Near the boundary, 
$$
A_{x} = A_{0} + A_{2} {r^2\over2}+\dots
$$
A short computation then verifies that the conductivity is given by,
$$
\sigma(\w) = {\vev{J_{x}}\over\vev{E_{x}}} = -i{\vev{J_{x}}\over\w\vev{A_{x}}} = -i{A_{2}\over \w A_{0}}
$$
Note that, since we are solving a linear equation but only care about this ratio, the overall scale of $A_{x}$ is immaterial.  We can use this freedom to set $a_{0}=1$, which simplifies the numerical problem.

Notably, we can analytically determine the $\w$-dependence of $\sigma$ for large and small $\w$ via standard power-series analysis.  Importantly, the scaling in the superfluid phase to be independent of $\Delta$ and $M$.  At small frequency, we find,
\be
Im[\sigma(\omega \ll 1)] \propto \w^{-1}
\ee
while for large $\w$ we have,
\be
Re[\sigma(\omega \gg 1)]\propto \omega^{-1/3} 
~~~~~~~~
Im[\sigma(\omega \gg 1)] \propto \omega^{-1/3}
\ee
This last result unsurprisingly differs from the AdS case, where $ Re[\sigma(\omega \gg 1)]=1$.  Reassuringly, they both match numerical results presented below, a nice sanity check.

\section{Phases of a Schr\"odinger Superfluid}\label{Sec:BHPhases}

Thus armed, we now get down to the business of finding a superfluid state in our non-relativistic holographic CFT and exploring its phase diagram.  A priori, the phase space is fairly high-dimensional -- specifying a point involves fixing $\Delta$ and $M$ to fix the theory, then tuning $\mu_{Q}$, $T$ and $\Omega$ to sweep out the phase diagram.  For simplicity, we will begin by picking convenient values $\Delta=6/5$, $M=1/2$ and $\mu_{Q}=1/8$, then dial the background Mass density, $\Omega$.  This will reveal a zero-temperature quantum phase transition at a critical value $\Omega_{*}$.  We will then fix $\Omega$ and vary $\mu_{Q}$, which will drive the superconducting phase transition from 2$^{nd}$ to $1^{st}$ order.  

\subsection{Varying $\Omega$ and a Quantum Phase Transition?}
We begin by fixing $\Delta=6/5$ and $M=1/2$, then set $\mu_{Q}=1/8$ and vary $\Omega$ between 0 and 1.  The basic results are presented in Figures 1, 2 and 3, which plot the condensate $\vev{\CO(T)}$ as a function of temperature, as well as the AC conductivities $Re[\sigma(\w)]$ and $Im[\sigma(\w)]$, for $\Omega$ = ${1\over16}$, ${3\over8}$ and 1, respectively.  These results are discussed in detail below.

\vspace{0.2cm}
{~~~~ $\bf\bullet ~~ \Omega\ll\Omega_{*}$}\\
For very small $\Omega$, the geometry remains essentially AdS until very close to the boundary, so we expect most low-energy physics -- such as superfluid condensation -- to very closely track familiar AdS results.  This turns out to be almost correct, modulo a surprise we'll explore shortly.

\begin{figure}[!h]
\centerline{ \epsfig{figure=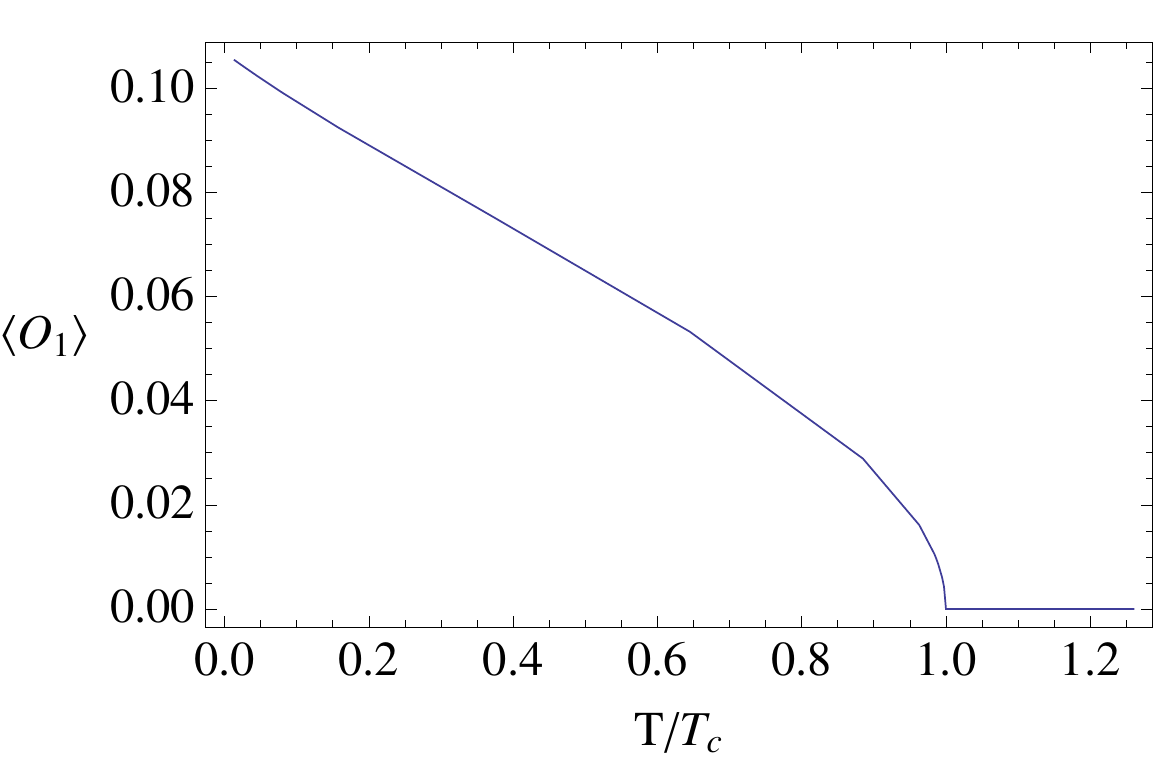, height=1.4 in}   \epsfig{figure=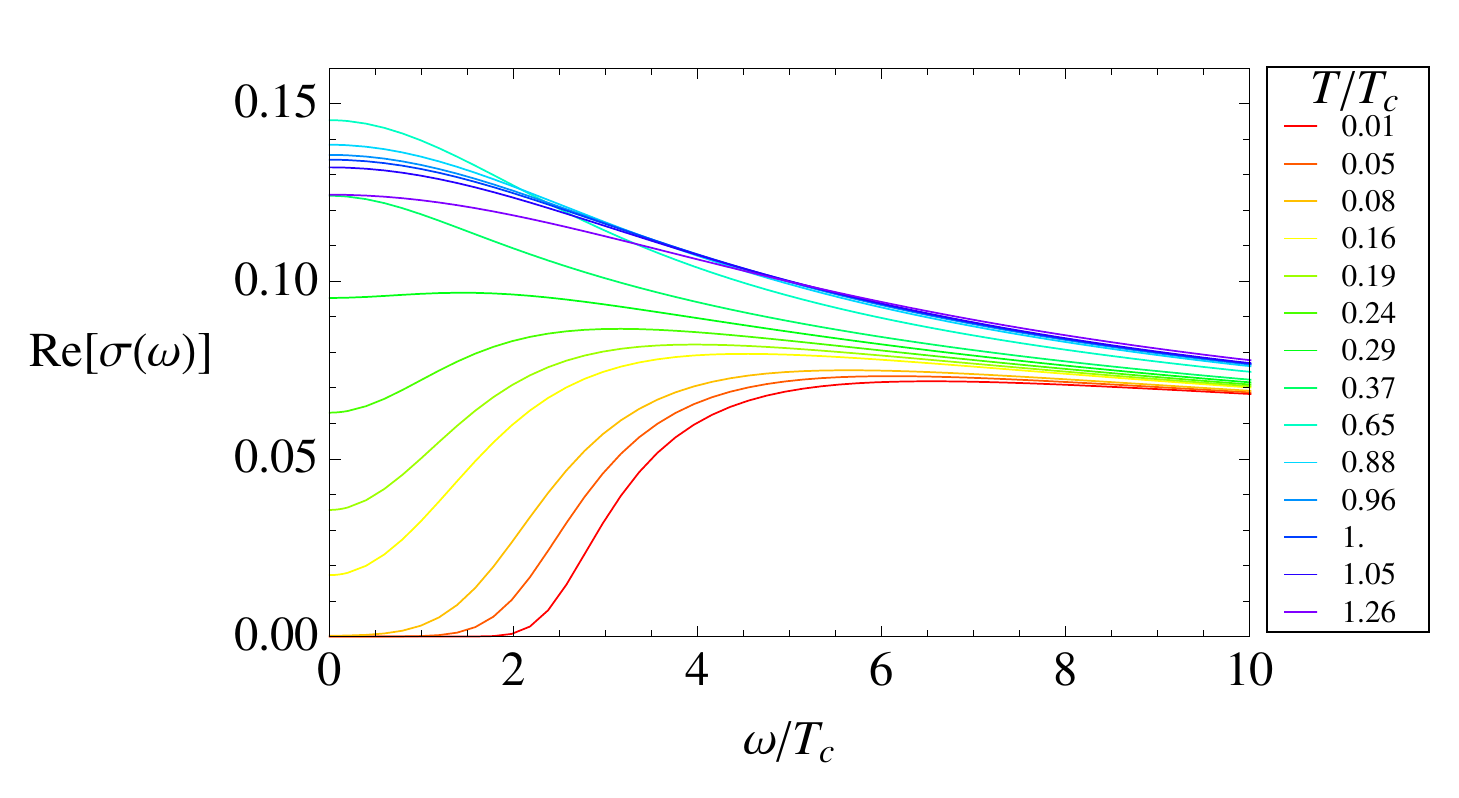, height=1.4 in}  \epsfig{figure=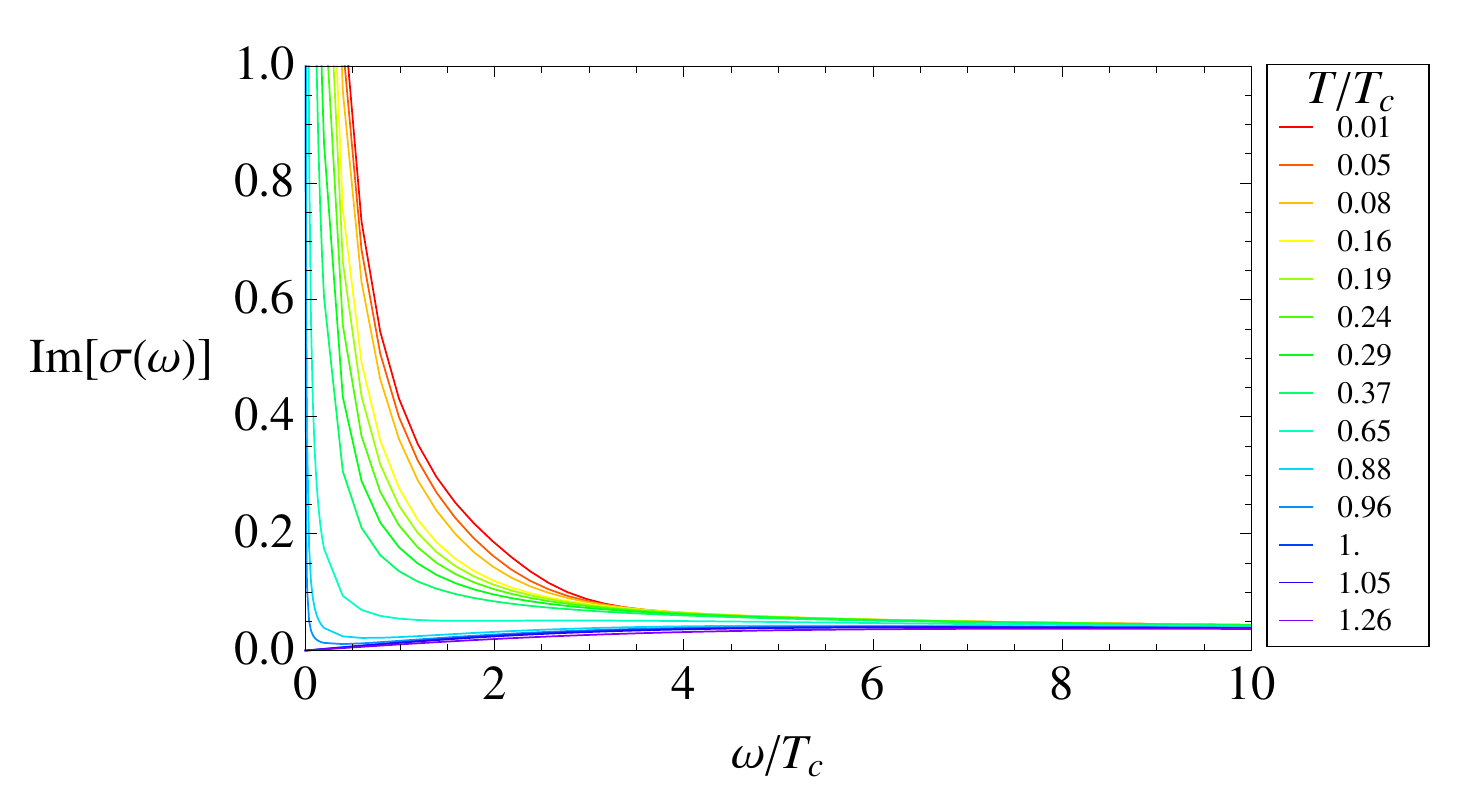, height=1.4 in} }
\caption{\em At small $\Omega$, the behavior of the superfluid is essentially the same as in AdS, with a $2^{nd}$ order mean-field phase transition at the onset of superconductivity at $T_{c}$, including the familiar gap-and-pole form in the AC conductivity, leaving us in a happy superfluid state at $T=0$.
Here, $(\mu_Q,\Omega)=(1/8,1/16)$, with $T_C=0.505$.
}
\label{FIG:TYPE1}
\end{figure}

Figure \ref{FIG:TYPE1}a shows the condensate as a function of temperature for the first conformal family, for $\Omega=1/16$.   As is clear by eye and can be checked precisely from the numerics, the resulting condensate turns on at $T=T_{c}$ with classic mean-field behavior ($\b_{c}=\half$) and grows as the temperature is lowered.   Figures \ref{FIG:TYPE1}b and \ref{FIG:TYPE1}c then show the real and imaginary parts of the AC conductivity for various temperatures indicated by color, from high (violet) to low (red).  These demonstrate the appearance of a superconducting state at $T_{c}$, with the gap growing as the temperature is lowered.   Note, too, that the conductivity in the superfluid phase has $Im[\sigma(\omega \rightarrow 0)] \sim 1/\omega$, while $Re[\sigma(\omega \rightarrow \infty)]\sim Im[\sigma(\omega \rightarrow \infty)] \sim \omega^{-1/3}$.  This scaling is expected on general grounds, so gives us confidence in our numerical results.

\begin{figure}[!h]
\centerline{ \epsfig{figure=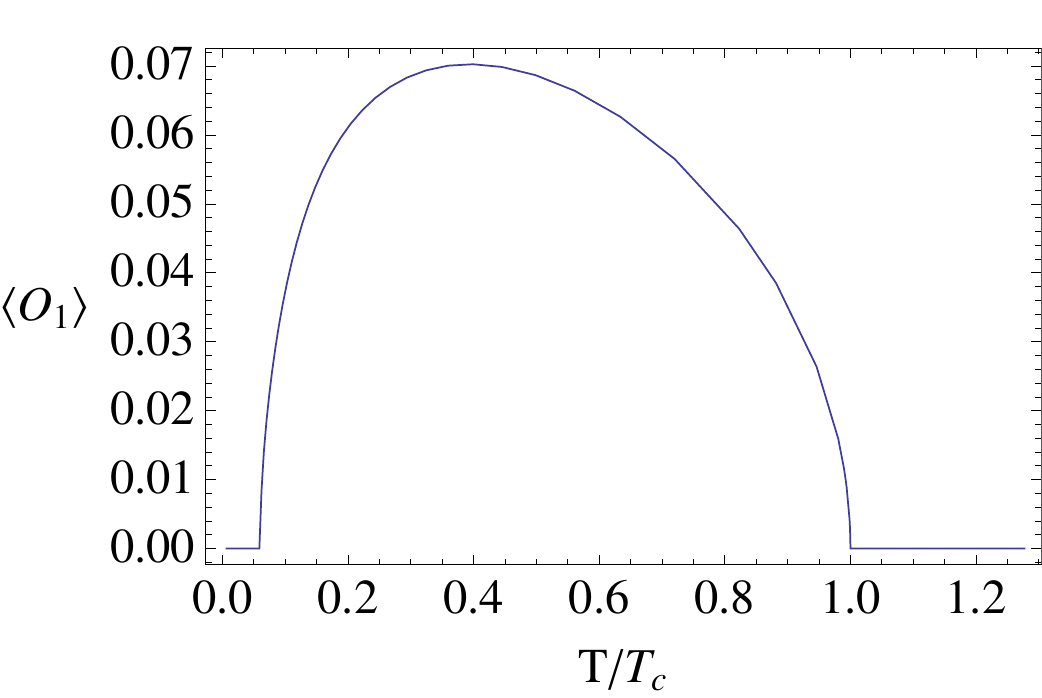, height=1.4 in}   \epsfig{figure=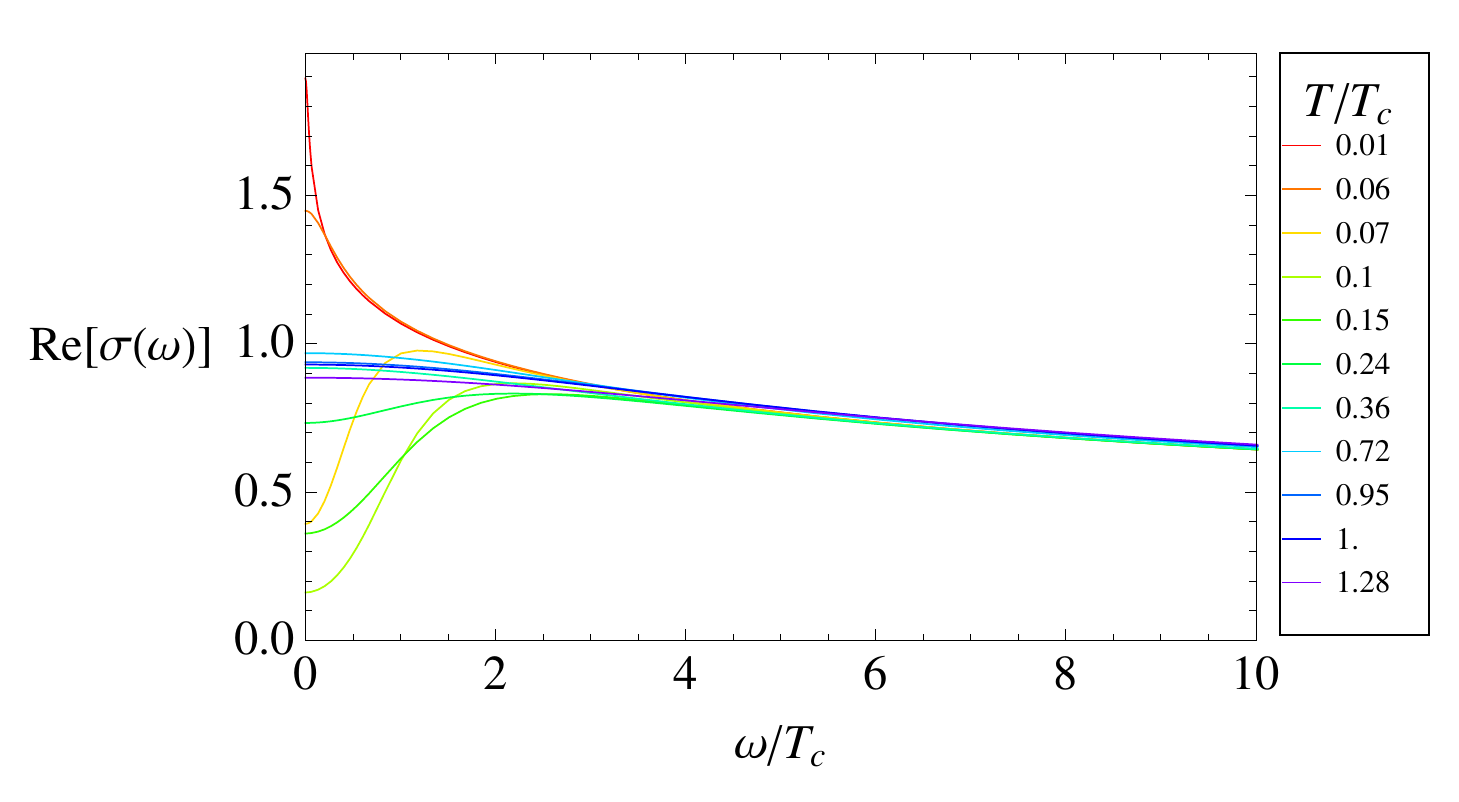, height=1.4 in}  \epsfig{figure=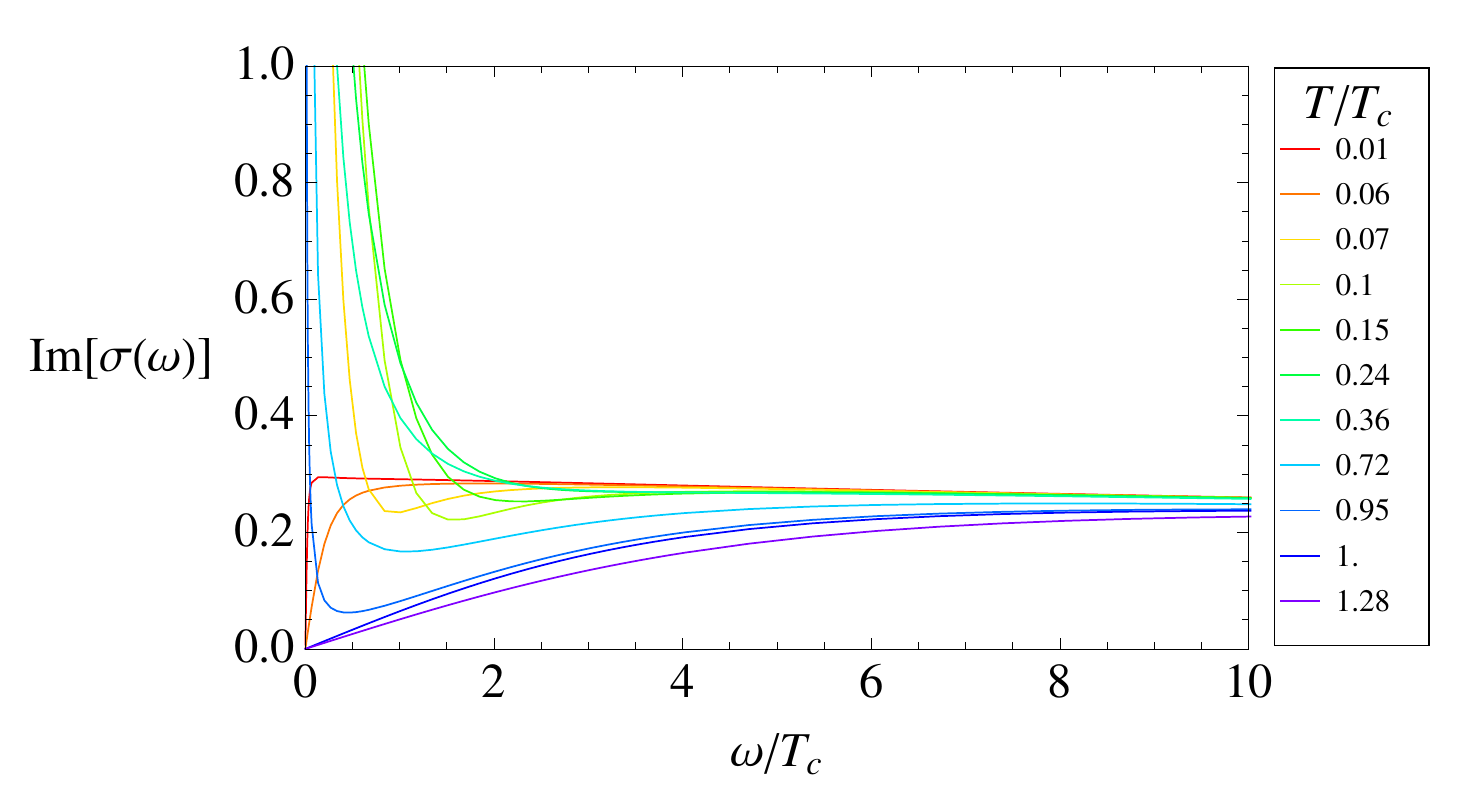, height=1.4 in} }
\caption{\em At large $\Omega$, in addition to the original transition to a superconducting state at $T_{c}$, the system now exhibits  reentrance of the normal phase at a new low-temperature $2^{nd}$ order transition at $T_{L}$, again with mean-field exponents.  Below $T_{L}$, $\sigma$ behaves like the normal gas.
Here $(\mu_Q,\Omega)=(1/8,1)$ with $T_C=0.149$ and $T_L=0.009$
}
\label{FIG:TYPE3}
\end{figure}

\newpage
{~~~~ $\bf\bullet ~~ \Omega\gg\Omega_{*}$}\\
As we increase the background number density, the story changes dramatically.  
Figure \ref{FIG:TYPE3} shows the same plots as Figure \ref{FIG:TYPE1} but with  $\Omega = 1$ rather than $\Omega=1/16$.  The most obvious difference is that the order parameter {\em vanishes} at sufficiently low temperature, $T\le T_{L}$, doing so again with mean-field behavior.  As is clear form the finite value of $Re[\s(0)]$,  the extreme low-temperature phase is again metallic.  

Consider now the behavior of the system at zero temperature as a function of the background number density, $\Omega$.  As $\Omega\to0$, the system is superconducting.  As $\Omega\to1$, the metallic phase is reentrant.  At some critical $\Omega_{*}$, then, the zero-temperature system appears to undergo a superconductor-metal quantum phase transition.

\begin{figure}[!h]
\centerline{ \epsfig{figure=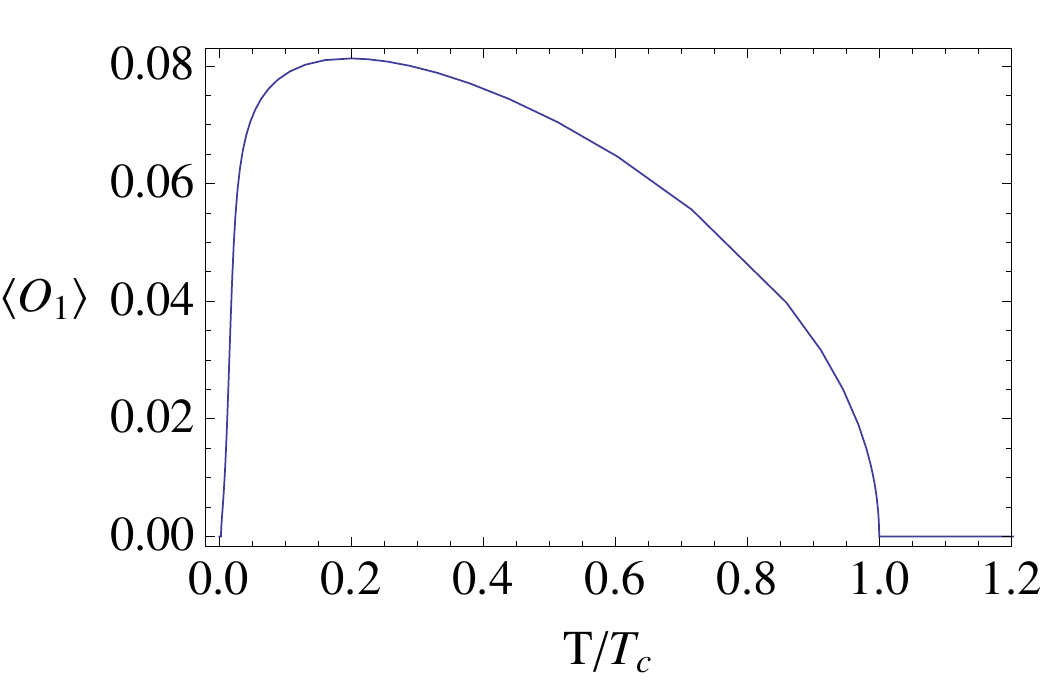, height=1.4 in}   \epsfig{figure=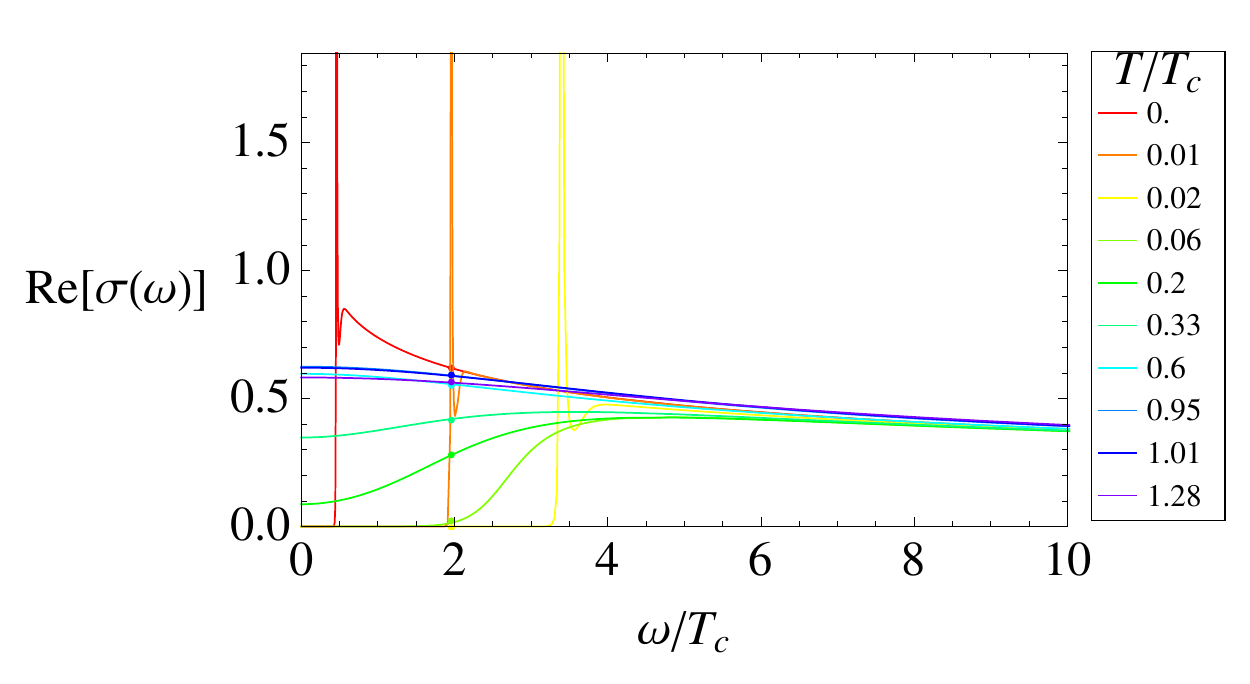, height=1.4 in}  \epsfig{figure=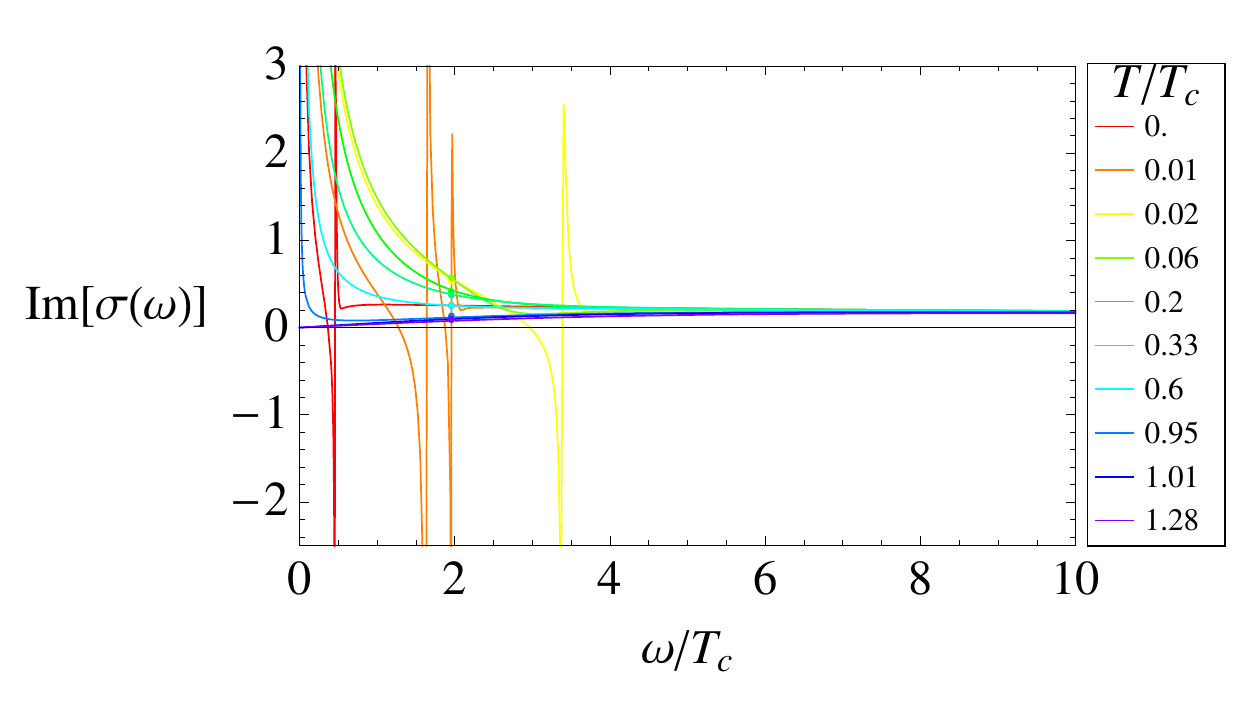, height=1.4 in} }
\caption{\em At intermediate $\Omega$, we again have a $2^{nd}$ order mean-field transition into a superfluid state at $T_{c}$.  At low temperatures, however, the system undergoes a non-mean-field transition to an apparently insulating state.
Here $(\mu_Q,\Omega)=(1/8,3/8)$ with $T_C=0.123$.}
\label{FIG:TYPE2}
\end{figure}

\vspace{0.2cm}
{~~~~ $\bf\bullet ~~ \Omega\to\Omega_{*}$}\\
It is tempting to try to determine what happens as we tune $\Omega$ towards this critical $\Omega_{*}$.  Figure \ref{FIG:TYPE2} shows the same system at $\Omega$ slightly above $\Omega_{*}$.
As before, there is a phase transition at $T_{c}$ with standard mean-field behavior.    The zero temperature behavior, however, differs dramatically from mean-field expectations; rather, at low temperature, the condensate decays exponentially, as does the superfluid density, while the normal density remains vanishing and the conductivity heavily suppressed at small but non-vanishing $\w$, suggesting that the $T=0$ state is not metallic.  It is tempting to read this as indicating a translationally-invariant insulating phase.

However, numerical results in this region should be taken with a sizeable grain of salt.  Indeed, at sufficiently low temperature, the numerics simply fail to converge.  More physically, in this regime, the probe approximation is becoming dangerously unreliable -- the matter field profiles which generate the required boundary values grow rapidly deep in the bulk (and in particular near the horizon) as we approach $T=0$ or $\Omega_{*}$.  Backreaction may thus qualitatively alter the low-temperature physics, either near the transition at $\Omega\sim\Omega_{*}$ or for sufficiently low $T$ at any $\Omega$.  

Indeed, it is entirely possible that the backreacted solution is re-entrant at any value of $\Omega$; our analysis is only reliable sufficiently far away from $T=0$.  
To unambiguously exclude re-entrance at small $\Omega$ as $T\to0$  requires including backreaction, which is beyond the scope of this paper.
Note, however, that the probe approximation shows no signs of inconsistency for $\Omega>\Omega_{*}$, so we can be quite confident that the system is definitely re-entrant at sufficiently large $\Omega$.

\subsection{High Temperature Condensates and the Free Energy}
\begin{figure}[!h]
\centerline{\epsfig{figure=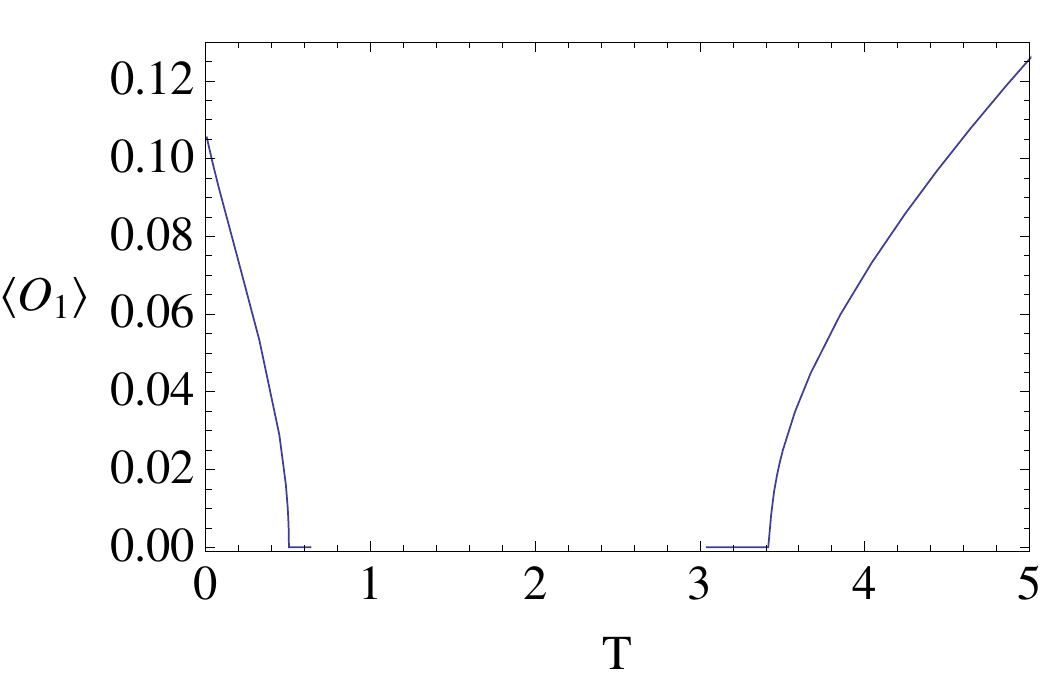, height=2 in} }
\caption{\em Surprisingly, there is another condensed phase at {\em high} temperatures.  Here $(\mu_Q,\Omega)=(1/8,1/16)$.\label{FIG:HighT}}
\label{type i}
\end{figure}

The surprise alluded to above involves the {\em high}-temperature limit.  Figure \ref{FIG:HighT} shows the same system but now extending to higher temperatures.  The surprise is the appearance of a {\em high temperature} condensate at $T\ge T_{H}$.  Troublingly,  the condensate appears to grow without bound as the temperature increases.\footnote{Such a high-temperature instability was predicted 
by Cremonesi {\em et al} \cite{Cremonesi:2009gy}\ whenever $\Delta\le4$.}

Before we panic, however, we should verify that this high-temperature condensate is in fact thermodynamically favored over the trivial vacuum.  Holographically, this means we computing the holographically renormalized on-shell action.  Unfortunately, in asymptotically Schr\"odinger spacetimes, holographically renormalizing the action is exceedingly complicated.   Happily, a simple strategy allows us to compute the difference in free energy between condensed and vacuum states without performing a full renormalization of the action.\!\footnote{We thank Nabil Iqbal for illuminating discussions on this topic, and refer the reader to \cite{Nabil},  in which this approach is further developed.}

The basic idea goes as follows.  Generally, specifying the non-normalizable (source) mode $\phi_{1}$ of the bulk scalar determines the normalizable (response) mode, $\phi_{2}$.  Smoothly varying the source thus traces out a curve $\phi_{2}(\phi_{1})$ in the ($\phi_{1}$, $\phi_{2}$) plane.    Along this flow, we can ask how the free energy -- aka, the Euclidean action -- varies.  Given the properly renormalized action, the variation of the full bulk action takes the form,
$
\delta S_{eff}=\dots \delta \phi_i+\dots \delta A_i
$
where $\delta \phi_i$ and $\delta A_i$ are the variations of the bulk fields and the $\dots$ correspond to the bulk equations of motion.  So long as we satisfy the bulk equations of motion, this reduces to a simple boundary term,
$
\delta S=(\Delta_1-\Delta_2) \int_{\partial_M} \phi_2 \; \delta \phi_1 -2\int_{\partial_M} (\rho_M \; \delta \mu_Q+\rho_Q \; \delta M_{o})
$
Moreover, if we hold fixed the asymptotic values of $A_{i}$ (corresponding to fixing the values of the chemical potential $\mu_{Q}$=$A_{t}|_{\p}$ and the mass $M$=$A_{\xi}|_{\p}$), this further simplifies to,
$
\delta S=(\Delta_1-\Delta_2) \int_{\partial_M} \phi_2 \; \delta \phi_1
$
We can thus compute the relative free energy density ($\CF_{A}$$-\CF_{B}$) between any two states $A$ and $B$ connected by such a flow by integrating $\delta S$ along the flow,
\be
\CF_B-\CF_A =-T\int^B_A \frac{\delta S_E}{V_{D}}=-T(\Delta_1-\Delta_2)  \int^B_A  \phi_2 \; d \phi_1 
\label{eq:F0}
\ee
where $V_{D}$ is the volume of the boundary theory and the integral is performed along the flow specified above.  By construction, this agrees with what we would get by evaluating the fully holographically renormalized free energy for each solution and subtracting.  Happily, this allows us to compute the correct free energy without having to worry about the full holographic renormalization of the theory (for further comparison between holographic renormalization and our method, see \cite{Guica:2010sw},\cite{Ross:2009ar},\cite{Marolf:2006nd}).

Now consider the case of our holographic superfluid in alternate quantization, where $\phi_{1}$=$\vev{\CO}$ is the response and $\phi_{2}$=$\CJ$ is the source.  In this case, the curve $\phi_{1}(\phi_{2})$ is multi-valued over $\phi_{2}$=0, with one solution corresponding to the trivial vacuum, $\vev{\CO}$=0, and one to the nontrivial condensate, $\vev{\CO}$$\neq$0.  As outlined above, these two solutions are connected by a very specific flow in the ($\phi_{1},\,\phi_{2}$) plane.  To compute the properly renormalized relative free energy, then, all we must do is find this flow and integrate along it,
\be
\CF_C-\CF_N=-T(\Delta_1-\Delta_2)  \int^C_N  \phi_2 \; d \phi_1 
\label{eq:F}
\ee
where the integration is again along the flow defined above.  If this difference is negative, the condensate is thermodynamically favored.

\begin{figure}[!t]
\centerline{ \epsfig{figure=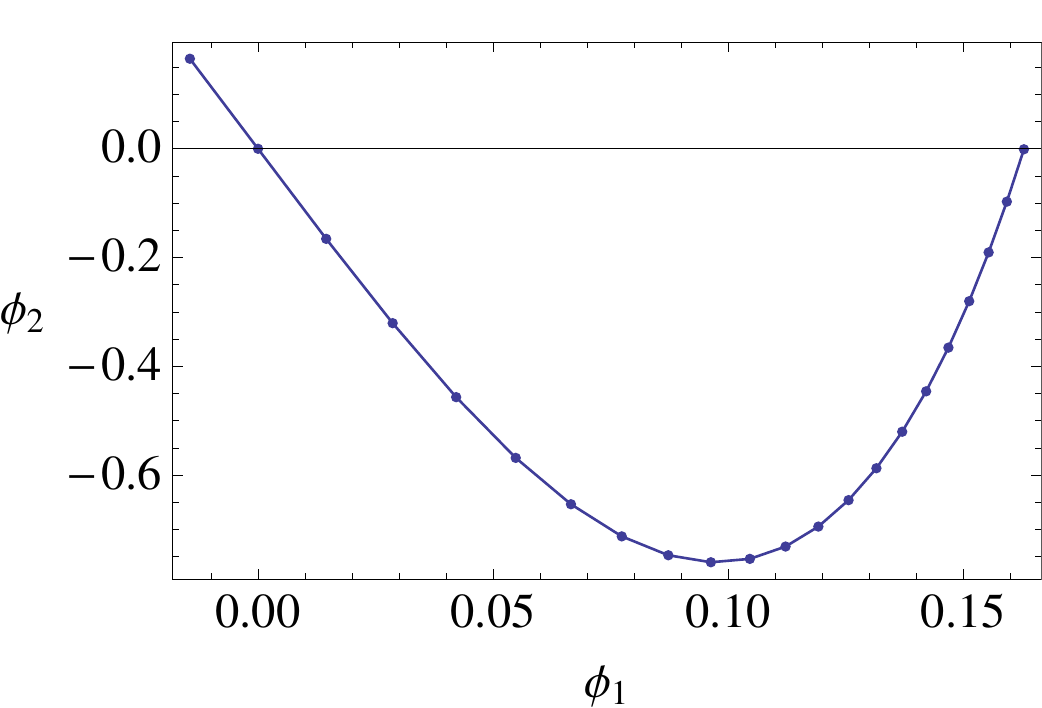, height=2in} ~~~~~~~~ \epsfig{figure=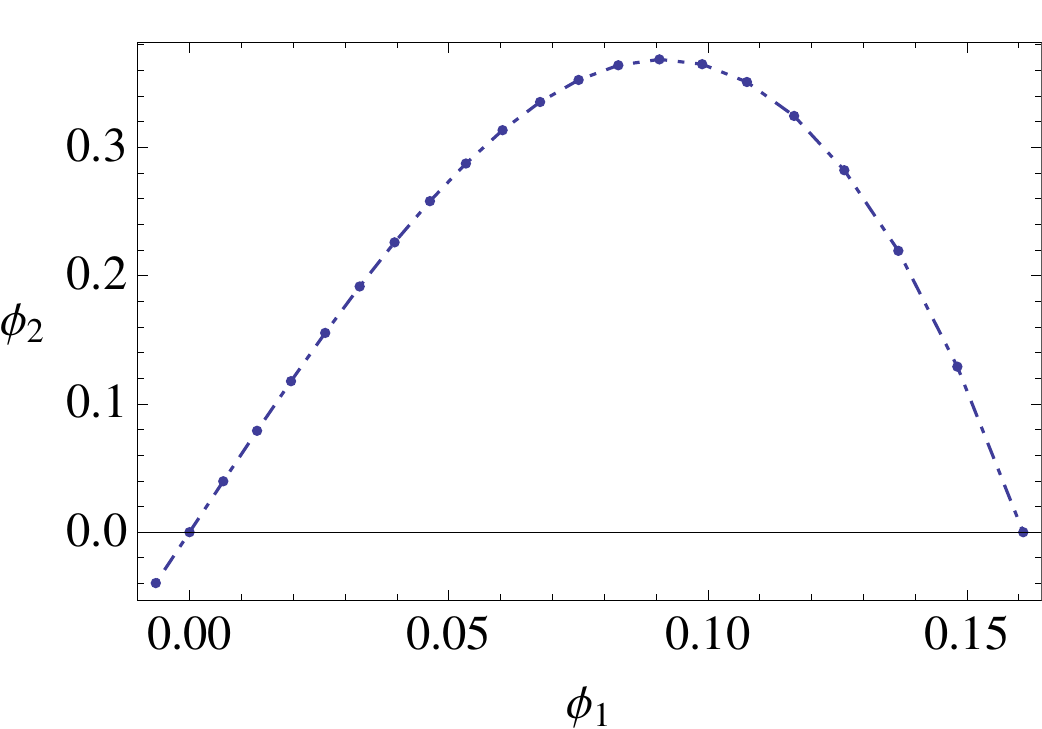,  height=2in}}
\caption{\em (a) Low T condensate has smaller free energy than non-condensed phase, $\CF_C-\CF_N <0$.  (b) High $T$ condensate has larger free energy than non-condensed phase, $\CF_C-\CF_N >0$.  One can determine the sign of $\CF_C-\CF_N $ from the orientation of the curve.  Here $(\mu_Q,\Omega)=(3/8,1/16)$.}
\label{lowhighTarea}
\end{figure}

Figure \ref{lowhighTarea} plots two such flows.  On the left we have a flow connecting the trivial vacuum ($\phi_{1}$=$\phi_{2}$=0) and a non-trivial vacuum ($\phi_{1}$$\neq$$0$, $\phi_{2}$$=$$0$) of the first conformal family in the low-temperature regime, with the flow indicated by the solid line and the direction of flow defining the direction of integration.  The area under the curve, corresponding to the free energy of the condensed state, is negative.  On the right is the analogous flow in the high-temperature regime -- here the free energy is positive.  We thus deduce that the low-temperature condensate is thermodynamically stable, while the high-temperature condensate is unstable, at least for this first conformal family.

\begin{figure}[!t]
\centerline{\epsfig{figure=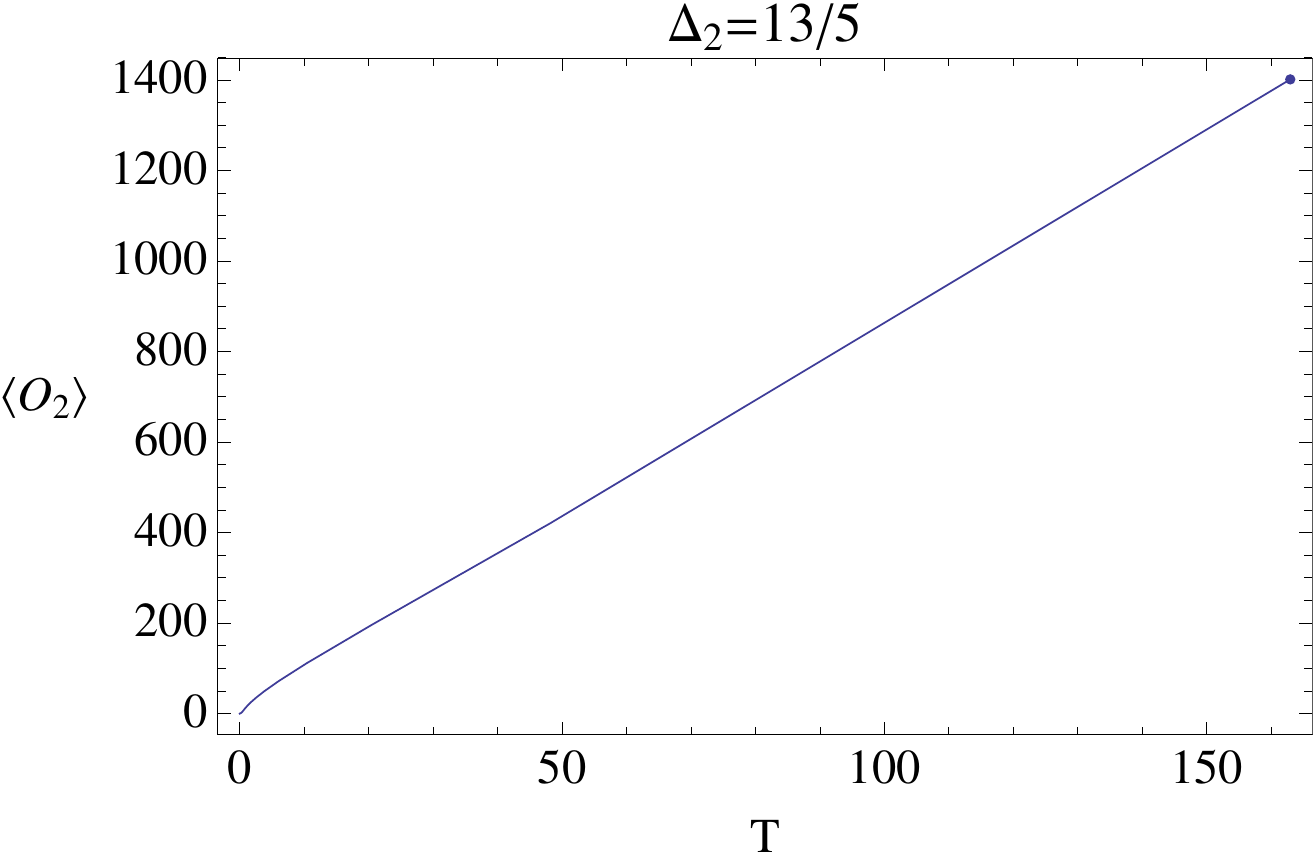,  height=2in} ~~~~~~~~ \epsfig{figure=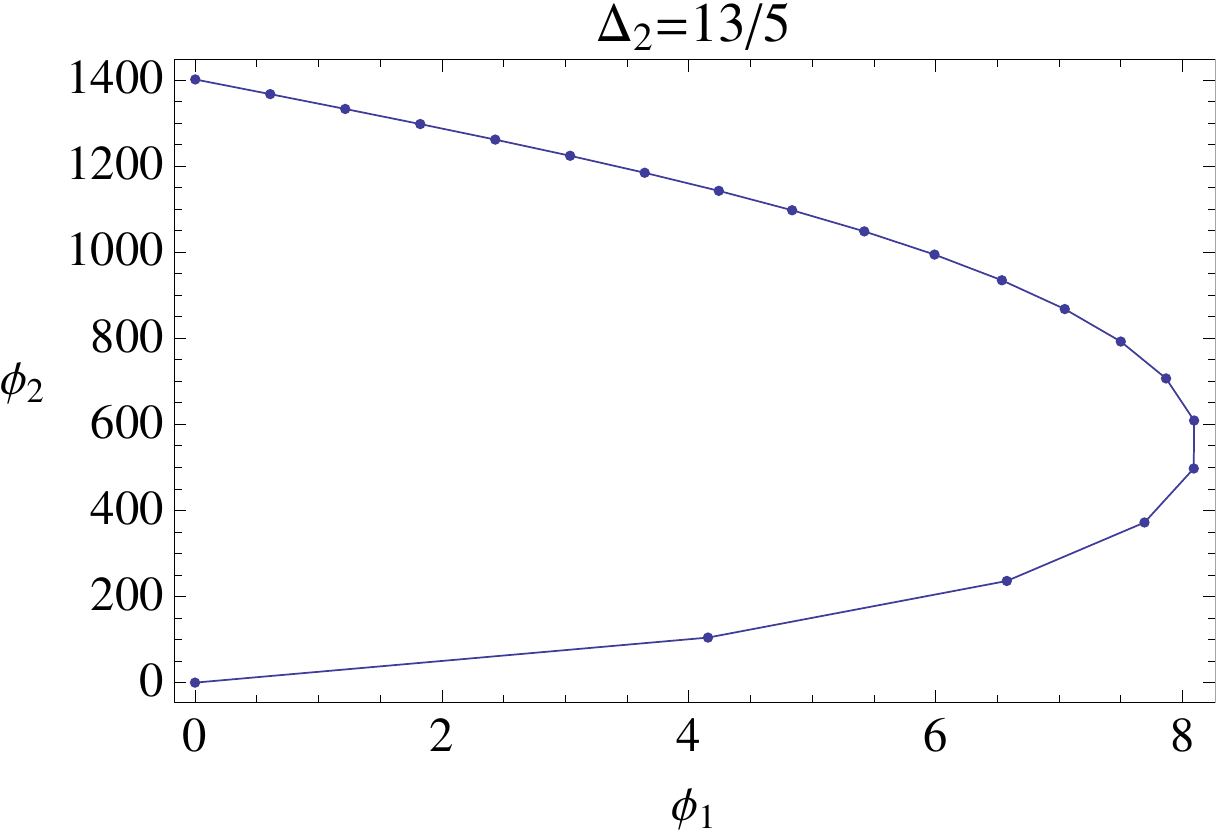,  height=2in} }
\caption{\em (a) Condensate as a function of T for the second conformal family with $\Delta_2=13/5$. (b) Typical flow at generic temperature, indicating a thermodynamic instability at every temperature.}
\label{FIG:2ndFam}\end{figure}
What about the second conformal family?  Figure \ref{FIG:2ndFam}a plots the condensate of the second family as a function of temperature.  Note that there is no separate low vs high temperature condensate, just a single continuous instability whose profile grows with temperature.  
Figure \ref{FIG:2ndFam}b then shows a typical flow at typical temperature.  Importantly, the enclosed area is negative for every temperature, indicating a thermodynamic instability even at arbitrarily high temperature.  This is why we quietly chose the first conformal family in Section 2.  It would be interesting to understand the meaning of this instability in detail.

\subsection{Varying $\mu_{Q}$ and a Multicritical Point}

\begin{figure}[!h]
\centerline{\epsfig{figure=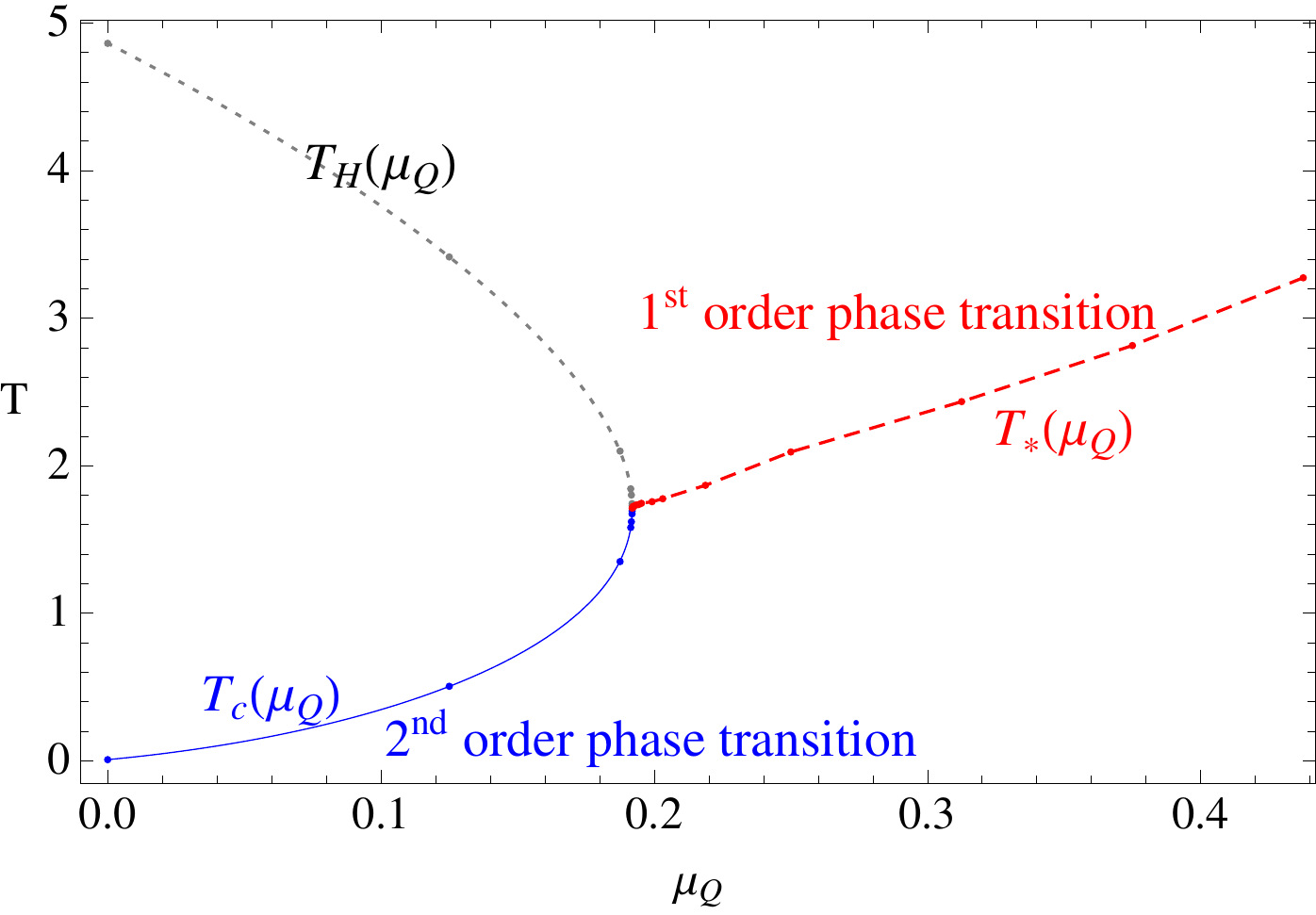, height=2.0 in} ~~ \epsfig{figure=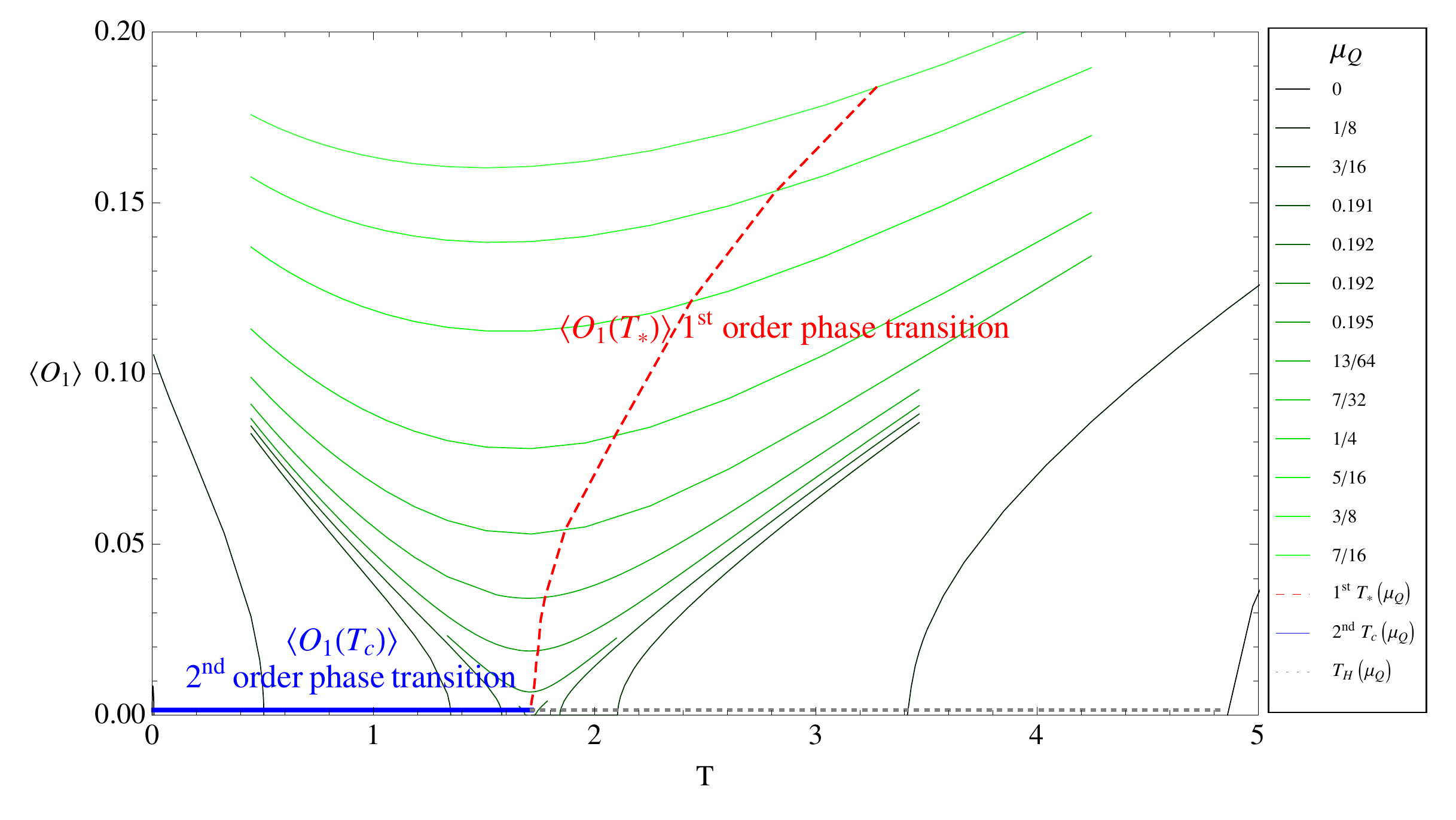,  height=2.0in} }
\caption{\em At $\mu_{Q}>\mu_{*}\sim0.192$, the transition goes 1$^{st}$ order.  Here, $\Omega$=${1 / 16}$.\label{FIG:FirstOrder}}
\label{type i}
\end{figure}

The thermodynamic instability of the high-temperature condensate leads to an important physical effect as we vary $\m_{Q}$. Figure \ref{FIG:FirstOrder}a plots $T_{c}(\mu_{Q})$ and $T_{H}(\mu_{Q})$, the critical temperatures for the low- and high-temperature condensates as a function of the chemical potential $\mu_{Q}$.  As we crank up $\mu_{t}$, holding all other parameters fixed, $T_{c}$ increases while $T_{H}$ decreases.  At a critical value, $\mu_{*}$, the two critical points merge; above $\mu_{*}$,  the condensate is non-zero for all temperatures.  This is clear from Figure \ref{FIG:FirstOrder}b, where we plot the order parameter as a function of temperature for values of the chemical potential above and below this critical $\mu_{*}$.

However, we have already checked that the condensed phase is thermodynamically disfavored at high temperatures.  For $\mu_{Q}>\mu_{*}$, then, there must be a critical temperature, $T_{*}$, above which the smoothly varying, non-vanishing condensate becomes thermodynamically disfavored.  This temperature is indicated by the red dashed curves in Figures \ref{FIG:FirstOrder}a and \ref{FIG:FirstOrder}b.

\begin{figure}[!h]
\centerline{ \epsfig{figure=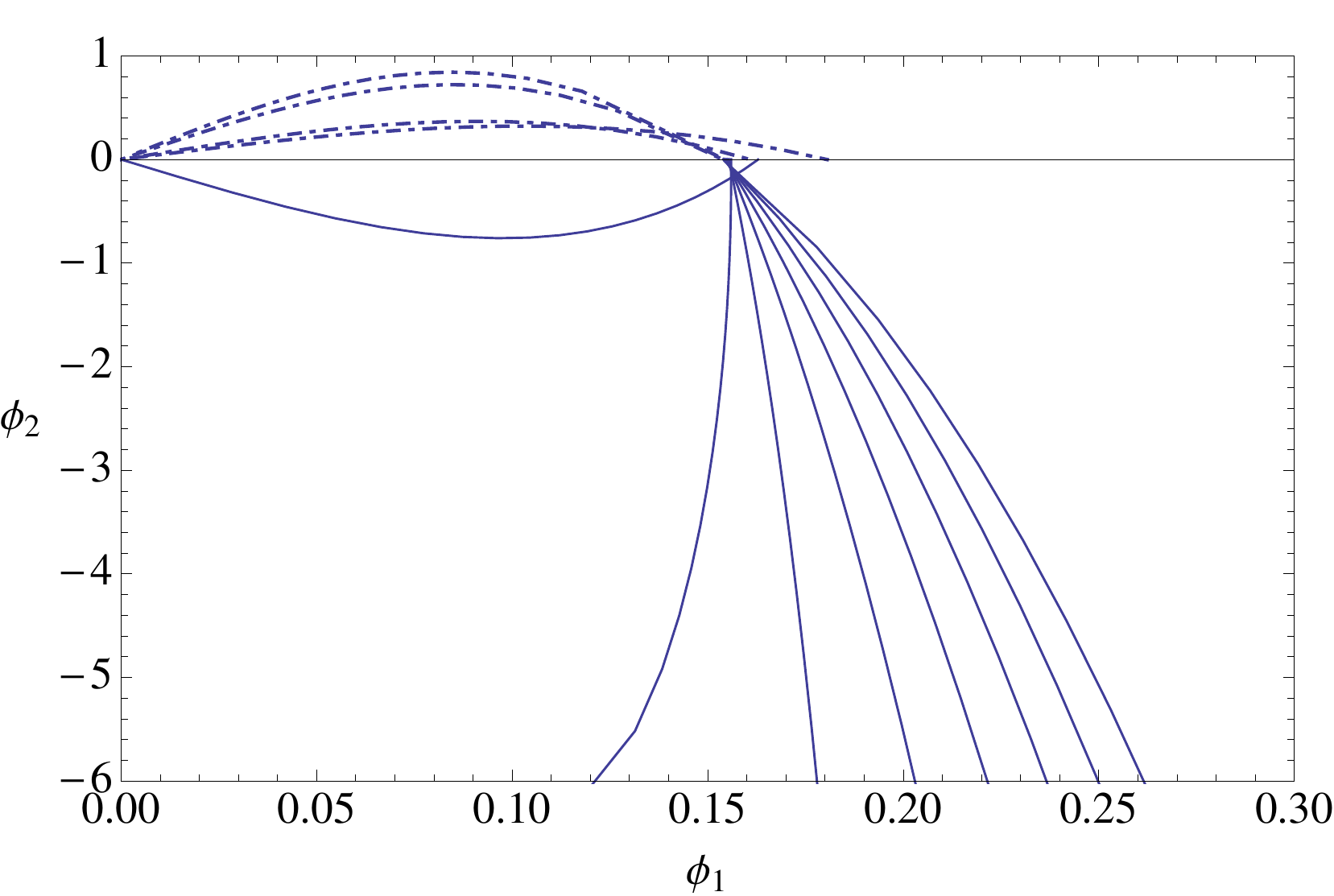, height=2in} ~~~~~~~~ \epsfig{figure=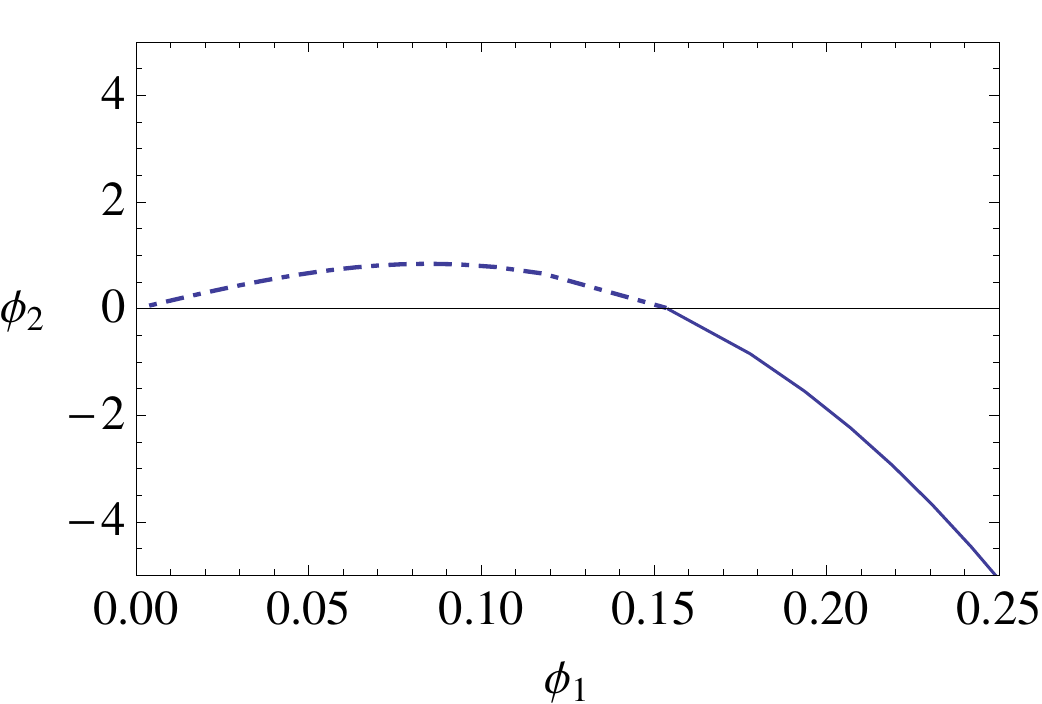,  height=2in}}
\caption{\em (a) Flow lines in the neighborhood of the critical temperature, for $\m_{t}=3$.  (b)  The flow switches direction discontinuously (lower solid to upper dashed curves) at a critical temperature $T_{*}$ indicated by the red dashed curve in 
Figure \ref{type i}, leading to a first order phase transition at $T_{*}$.  Here, $\mu_{Q}$=$3/8$ and $\Omega$=$1/16$.
\label{O1_O2_area}
} \end{figure}
We can verify this by computing the relative free energy of the condensed phase as we vary the temperature.  Figure \ref{O1_O2_area}a shows the flows associated to points to the left and right of the critical temperature where the low- and high-temperature instabilities meet (indicated by the red dashed curve in Figure \ref{FIG:FirstOrder}b).  Figure \ref{O1_O2_area}b focuses in on the immediate neighborhood of the transition temperature for $\mu_{Q}>\mu_{*}$.  For all temperatures below the critical temperature, the flows go below the horizontal axis, corresponding to a negative free energy and a thermodynamically stable condensate.  For all temperatures above the critical temperature, the flows go above the horizontal axis, so the condensate is thermodynamically disfavored at high temperatures. 
Indeed, while the value of the condensate is non-vanishing and in fact completely smooth as we flow through $T_{*}$, the path which carries us from the trivial state to the condensate changes discontinuously as we pass through $T_{*}$.   As a result, the integrated area -- and thus the free energy -- also changes discontinuously at $T_{*}$.   Moreover, as we take $\mu_{Q}\to\mu_{*}$, the value of the condensate at the transition goes to zero, $\vev{\CO(T_{*})}\to0$; this ensures that the latent heat of the transition goes to zero at the multicritical point where the transition switches from 1$^{st}$ to 2$^{nd}$ order, as expected on general grounds. 

The upshot of all of the above is that as we raise $\mu_{Q}$, the phase transition from high-temperature metal to low-temperature superfluid switches from 2$^{nd}$ order to 1$^{st}$, with the transition occurring at a multicritical point where the low- and high-temperature superfluid phases collide.
Near the phase transition boundaries, including the multicritical point, the order parameter scales with simple mean-field exponents. More precisely, near the finite temperature 2$^{nd}$ order phase transition,  $\langle \CO \rangle \sim (T_{c}-T)^{1/2}$, while near the 1$^{st} $ order phase transition boundary when $\mu_Q>\mu_*$ the condensate jumps discontinuously at $T_*$, with $\langle \CO(T_{*}) \rangle \sim (\mu_Q-\mu_*)^{1/2}$. This can be succinctly encoded in a simple mean-field free energy, $F(\varphi)= \frac{1}{2}c_2 (T-T_c(\mu_Q))\varphi^2+\frac{1}{4}c_4(\mu_*-\mu_Q) \varphi^4+ \frac{1}{6}c_6 \varphi^6$, with $\varphi\sim \langle \CO \rangle$ and with coefficients $c_{2},c_{4},c_{6}>0$.

\subsection{Setting $\mu_{Q}=0$ and the Persistence of Condensates}

Playing with $\mu_{Q}$ raises another interesting point.  Fundamental to our construction is that the scalar operator carries a charge $q$ under a global symmetry of the boundary theory.  $\mu_{Q}$ tells us the energy cost for adding a unit of this charge to the system.  In AdS, superfluid condensation is often induced by tuning $\mu_{Q}$ beyond a threshold.  Is this also necessary in the non-relativistic case?  

\begin{figure}[h]
\centerline{\epsfig{figure=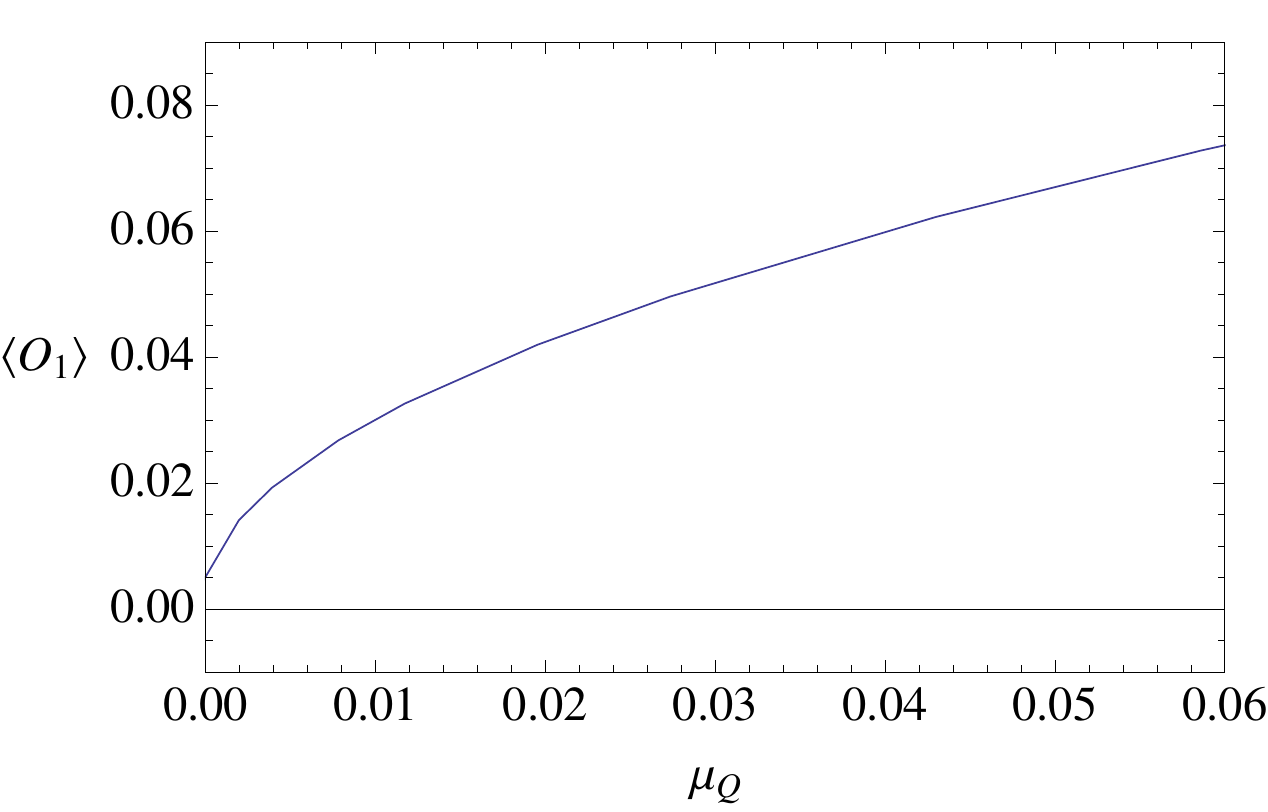, height=2 in} ~~~~~~~~ \epsfig{figure=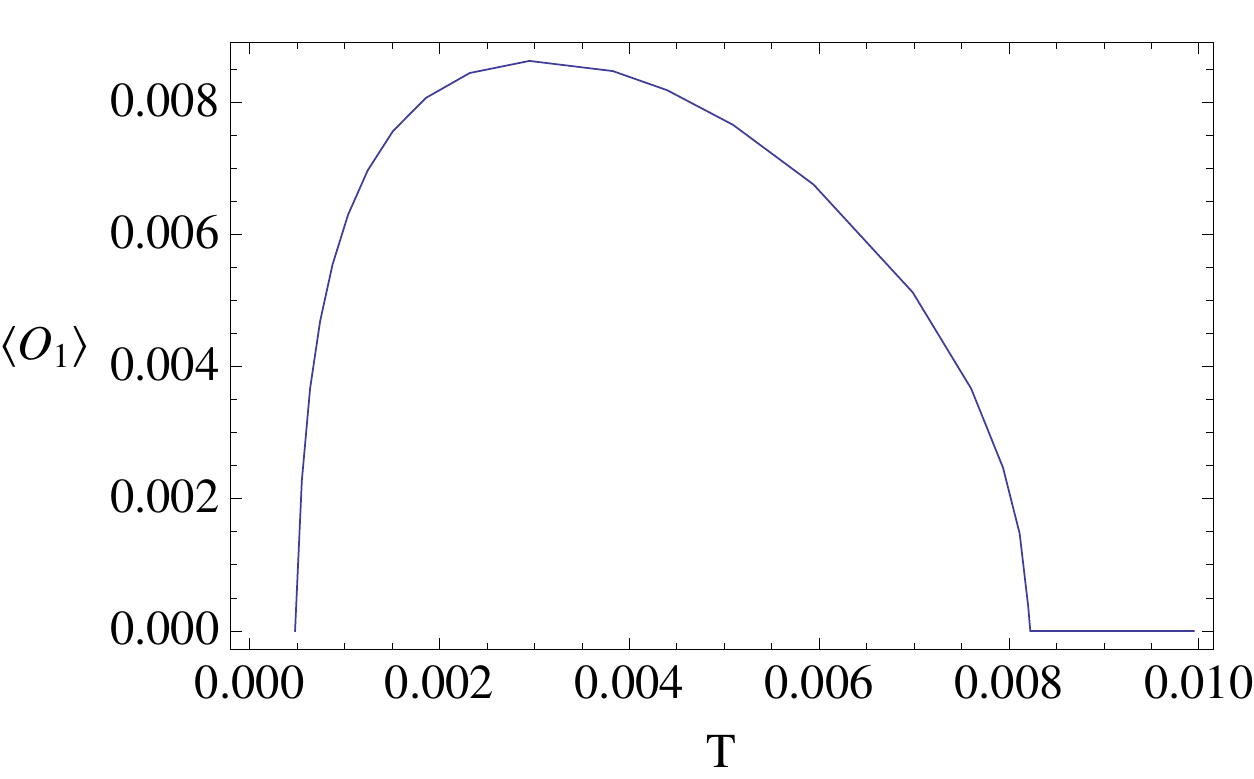, height=2 in}   }
\caption{\em (a) Fixing $T\ll1$ and varying $\mu_{t}$.  (b) Fixing $\m_{t}=0$ and varying $T$. Here, $\Omega$=$1/16$.\label{FIG:Pairing}}
\end{figure}
By varying $\mu_{Q}$ to zero while holding all other parameters fixed (see Figure \ref{FIG:Pairing}a), we see that condensation persists even at vanishing $\mu_{Q}$.  Indeed, by plotting $\vev{\CO(T)}$ for $\mu_{Q}=0$ (see Figure \ref{FIG:Pairing}b), we see that the form of this curve is quite similar to the large-$\Omega$ case studied above with $\mu_{Q}\neq0$, modulo an overall scaling of the condensate.  It is tempting to speculate that this indicates two distinct pairing mechanisms, one involving the charge and one involving the Mass eigenvalue alone.  It would be interesting to explore this point further.

\section{Conclusions and Open Questions}\label{Sec:Conclusions}

In this paper we have constructed a toy model of superfluid states in holographic NRCFTs and studied the resulting phase diagram, finding several unanticipated features.  First, as we lower the temperature in the disordered metallic state, the system generally undergoes a phase transition to a superfluid state.  At small (and even vanishing) chemical potential, this transition is $2^{nd}$ order with mean-field exponents; at large chemical potential, however, the transition runs strongly $1^{st}$ order.  Secondly, for large background mass density, the superfluid state only appears in a finite temperature window, with the metallic state reentering at sufficiently low temperature.  Finally, at zero temperature, the reentrance of the metallic phase leads to an apparent quantum phase transition from superconducting to metallic as the background mass density is varied.

Several features of our results deserve further scrutiny.  First, our low temperature results derive from a probe analysis which is not valid at zero temperature -- indeed, as we push the temperature to zero near or below the putative quantum phase transition at $\Omega_{*}$, the bulk profiles of various fields grow rapidly, diverging as we approach zero temperature.  To be sure, we checked the consistency of the probe approximation in each calculation presented above. 
However, it is entirely possible that various of our results could change qualitatively when we include backreaction.  To nail down the $T=0$ physics, we must incorporate backreaction.  

Relatedly, we have tacitly assumed that the neutral black hole geometry is the dominant saddle at $T=0$.  However, for a variety of reasons including the strange thermodynamics of this black hole, this seems unlikely to be the case.  It is tempting to speculate that the low-temperature phase is dominated by a Schr\"odinger soliton analogous to the AdS soliton which dominates the relativistic case a la Hawking-Page.  Indeed, a simple such Schr\"odinger soliton solution is known, and we repeat the above analysis for this geometry in an Appendix.  However, the black hole and the soliton enjoy incommensurate asymptotic periodicity conditions, so cannot contribute to the same ensembles.  Understanding the true low-temperature ground state of the neutral black hole, even in the absence of any charge density in the system, is of considerable interest.

To this end, it is worth emphasizing that the basic trouble with the thermodynamics of this -- and indeed all known -- asymptotically Schr\"odinger black holes is the light-cone relation between the near-horizon killing vector $\p_{\tau}$ and the asymptotic timelike killing vector, $\p_{t}$.  In most constructions, this follows from the structure of the salient solution generating technique.  The challenge, then, is to build solutions which do not flow to $AdS$ black holes near the horizon.

Meanwhile, it's important to keep in mind that the simple Abelian-Higgs theory we study is an extremely stripped down toy model for which we do not have an explicit charged black hole solution.  In the few examples where such a solution is known \cite{Adams:2009dm, Imeroni:2009cs}, the matter sectors are considerably more complicated, which is why we worked with the toy model at hand as a first step.  It would be interesting to repeat our analysis in one of these more elaborate systems to disentangle the peculiarities of our toy model form general features of non-relativistic holographic superfluids.

Finally, several intriguing features of this system still need interpretation.  Why does the second conformal family have a thermodynamically dominant high-temperature instability, what does that instability signal, and is there a simple low-energy way to see that this family is disfavored?  Our conductivity calculations reveal a number of quasiparticle peaks with rather peculiar behavior, particularly near the critical point at $\Omega_{*}$ -- are these artifacts of the probe approximation, or do they signal real physics, and if so, what are they telling us?  Does a proper holographic renormalization of the full system (which remains an open problem) alter any of our results, and if so how?  What about the pairing mechanism -- what is the fermion spectral function in these systems, and can we correlate pairing of probe fermions with the condensation seen above?  We hope to return to various of these questions in the future.


\section*{Acknowledgments}

We would like to thank 
K.~Balasubramanian, 
O.~DeWolfe, 
N.~Iqbal, 
S.~Kachru, 
H.~Liu, 
R.~Mahajan,
J.~McGreevy, 
Y.~Nishida, 
K.~Rajagopal,
and
D.~Vegh 
for valuable discussions.
AA thanks the Aspen Center for Physics and the Stanford Institute for Theoretical Physics for hospitality during the completion of this work.  
The work of AA is supported in part by funds provided by the U.S. Department of Energy (D.O.E.) under cooperative research agreement DE-FG0205ER41360.

\appendix

\section{Superfluids in a Schr\"odinger Soliton}
\label{Sec:SolPhases}


It is entertaining to apply the approach developed above to a slightly different geometry, the so-called Schr\"odinger soliton:\cite{Mann:2009xk}:
\be\label{EQ:Schrsoliton}
ds^2_{soliton,Str} = 
\(-f_s + {(f_s-1)^{2}\over 4(K_s-1)}\){dt^{2}\over K_s r^{4}}
+ 
{1+f_s \over r^{2} K_s} dt \, d\xi 
+{K_s-1\over K_s}d\xi^{2}
+ {d\vec{x}^{2} \over r^{2}}
+ {dr^{2}  \over f_s \, r^{2}}\,.
\ee
where $K_s=1-r^2\Omega^2$ and $f_s=1-r^4/r_s^4$, with  $r_{s}$ controlling both the gap and the radius of the $\xi$-direction, $L_{\xi}$=$\pi\over\Omega r_{s}$.  Here, the radial direction is cut off smoothly by a spacelike circle shrinking rather than by a black hole horizon -- and indeed this solution was obtained by double wick rotating the Schr\"odinger black hole with compact $\xi$-direction (which is spacelike near the horizon).\footnote{It is tempting to identify this solution as a low-temperature confined phase of the Schr\"odinger black hole studied above, analogous to the Hawking-Page transition from AdS black hole to AdS-soliton. 
However, this is not  correct: regularity of the euclidean solutions imposes incompatable periodicity conditions.  More precisely, for the black hole, smoothness of the global euclidean geometry and compactness of the direction dual to the Mass operator require the periodicities 
$i t_b \equiv i t_b+ n/T$ and $\xi_b \equiv \xi_b+ n \mu/T + w L_\xi.$
For the soliton, on the other hand, we need 
$i t_s\equiv i t_s - i n/T_s$ and $\xi_s \equiv \xi_s + i n \mu/T_s+ w L_\xi.$
The inequivalence  of these conditions (note in particular the extra factor of $i$ in $T_{s}$) tells us that these solutions correspond to inequivalent ensembles.
}

As in the black hole case, a probe superfluid in the soliton geometry is characterized by five parameters: two define the theory (the dimension $\Delta$ and Mass $M$), two are properties of the background which fix thermodynamics quantities (the mass density $\Omega$ and mass gap of the soliton, $m_{G}$=$1\over L_{\xi}$) and finally the $U(1)$-charge chemical potential $\mu_{Q}$ determined by the non-normalizable mode of the bulk gauge field $A_{t}$.  In the remainder of this appendix we briefly summarize the results.

\subsection{Varying $\mu_{Q}$} 
\begin{figure}[!h]
\centerline{\epsfig{figure=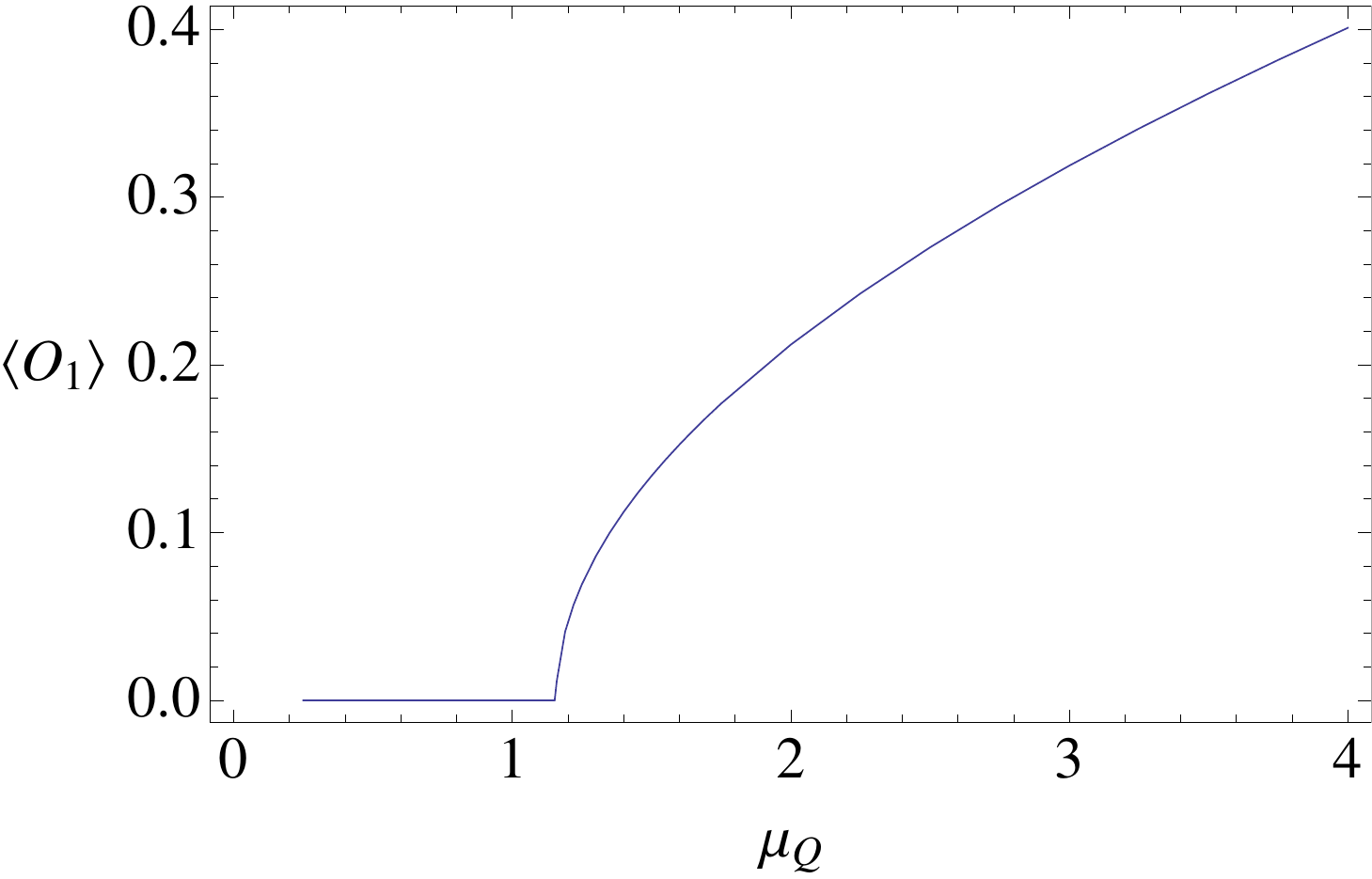, height=1.4 in}  \epsfig{figure=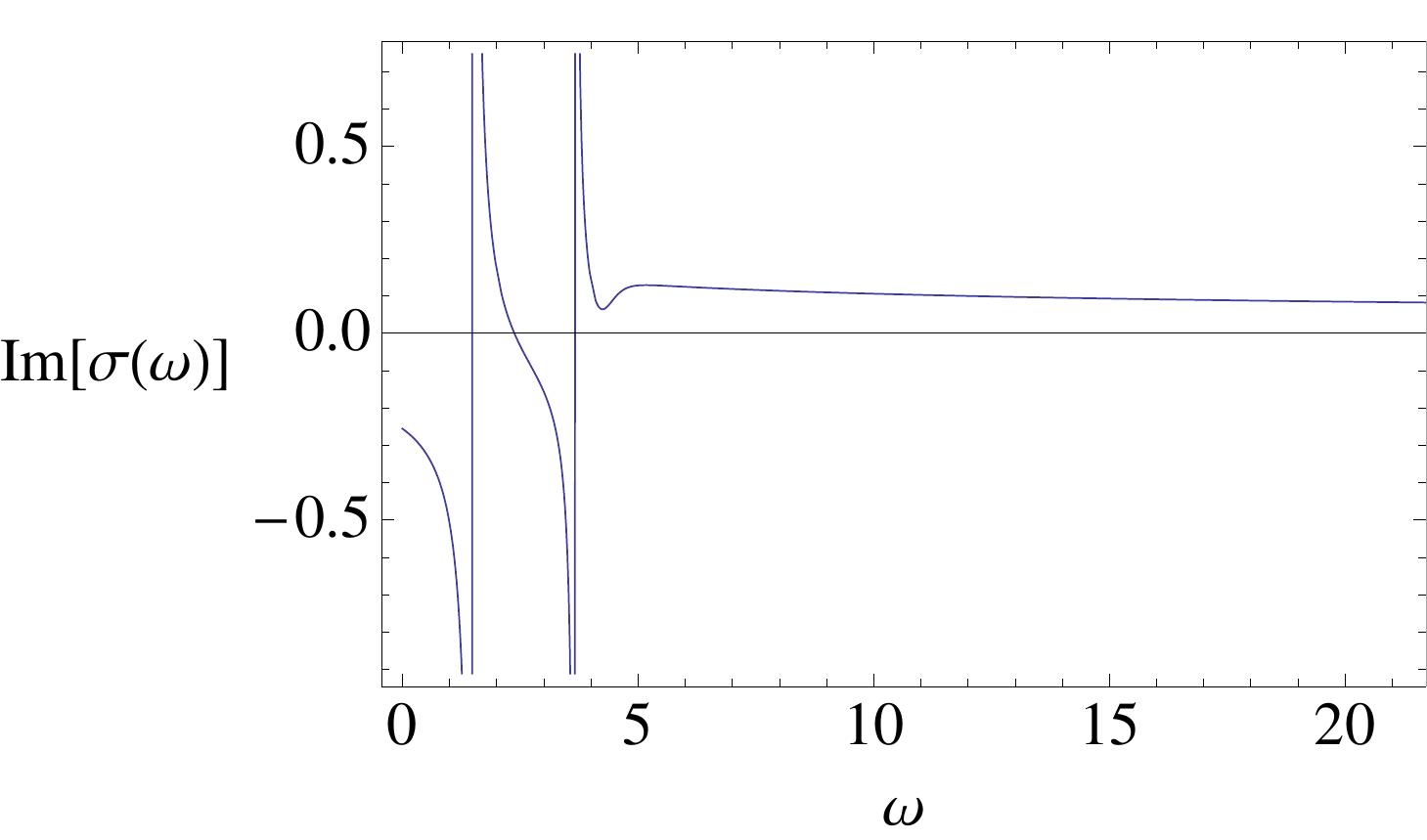, height=1.4 in} \epsfig{figure=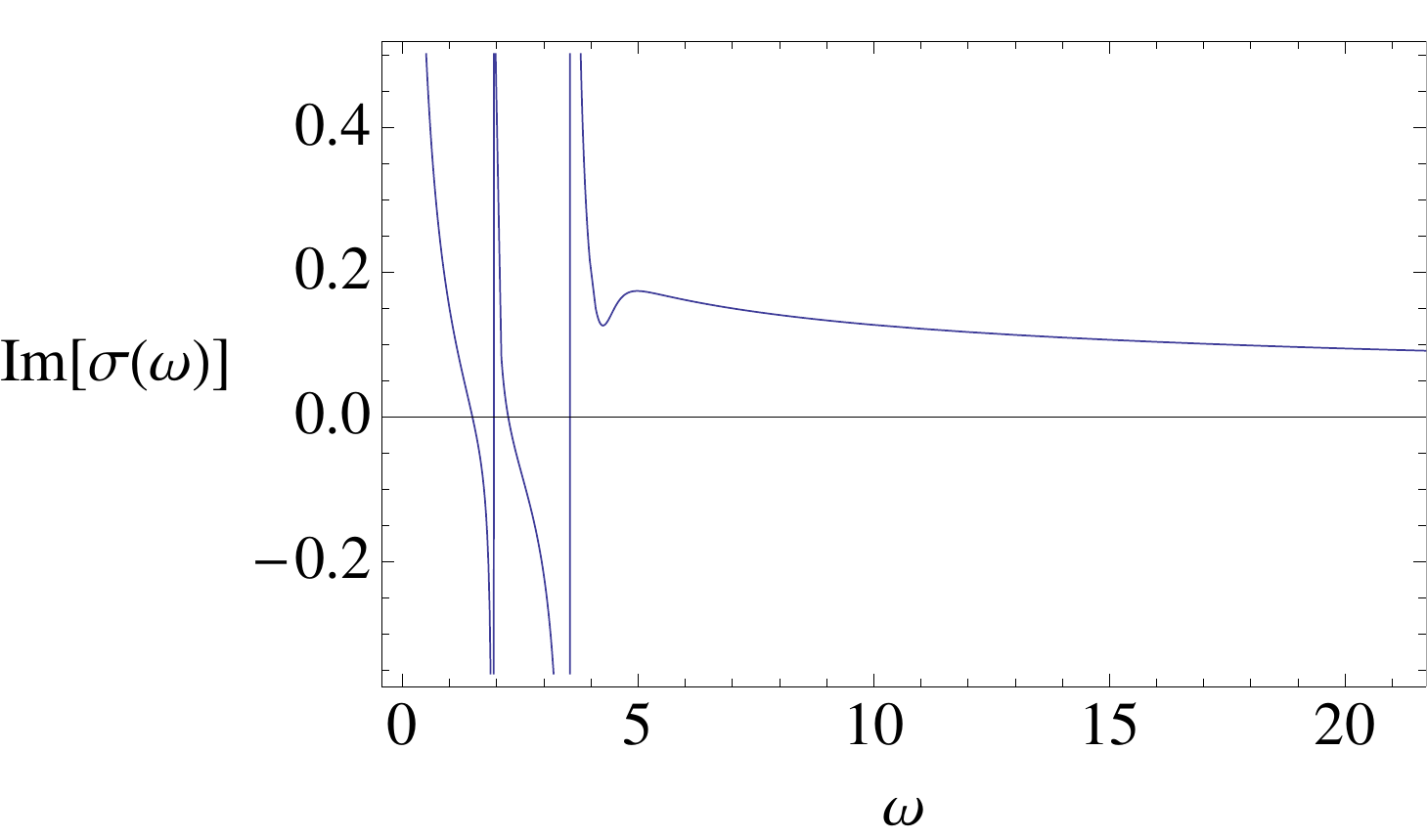, height=1.4 in} }
\caption{\em (a) We find a $2^{nd}$ order phase transition at $\mu_c=1.17$.  (b) When $\mu<\mu_c$, $Im[\sigma(\omega \to 0)]$ is finite, indicating an insulating phase.  (c) For $\mu>\mu_c$, we find a superconducting $1\over\w$ pole, as well as two gapped poles at finite $\w$.}
\label{fig11}
\end{figure}
Figure \ref{fig11}(a) shows the condensate as a function of the chemical potential $\mu_{Q}$, revealing a critical minimum value $\mu_{c}$ at which the system undergoes a 2$^{nd}$ order transition with mean-field exponent.  For $\mu<\mu_{c}$, $Im[\sigma(\omega\to0)]$ $\to$ finite, indicating a translationally invariant insulating phase.  For $\mu_{Q}>\mu_{c}$, by contrast, $Im[\sigma(\omega\to0)]\sim{1\over\w}$, indicating  superconductivity.   We would thus appear to find a $2^{nd}$ order insulator-superconductor quantum phase transition by varying $\mu_Q$.
However, in addition to this $\w\to0$ pole, we find two more mysterious poles at $\w_{1}$ and $\w_{2}$ separated by a finite gap.  This is reminiscent of the paired poles we found in the black hole system at intermediate values of the background density near the critical point at $\Omega_{*}$.  This transition also recalls the AdS transitions studied in  \cite{Nishioka:2009zj}.

\subsection{Varying $\Omega$}

\begin{figure}[!h]
\centerline{ (a)\epsfig{figure=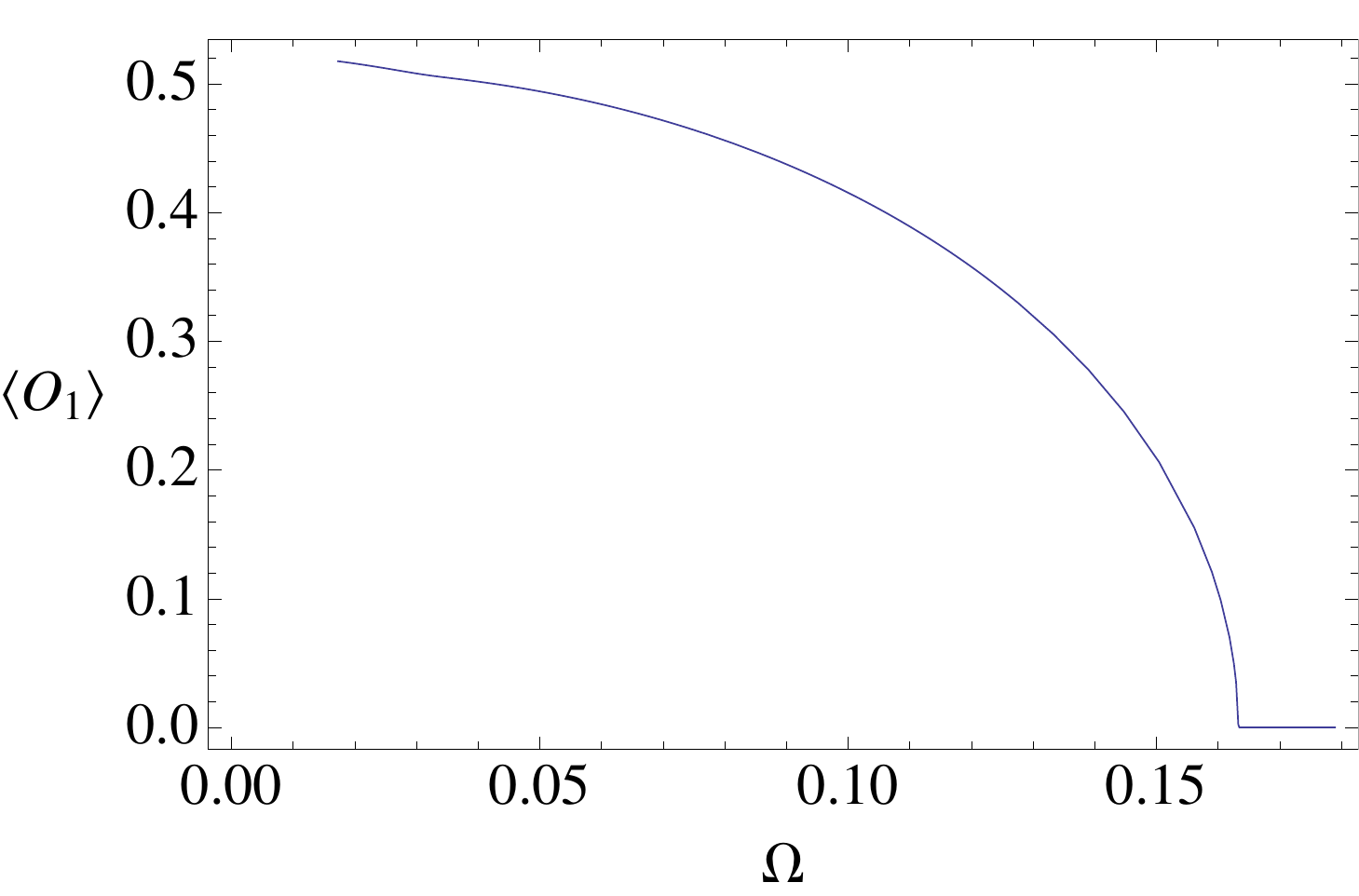, height=1.4 in}  (b)\epsfig{figure=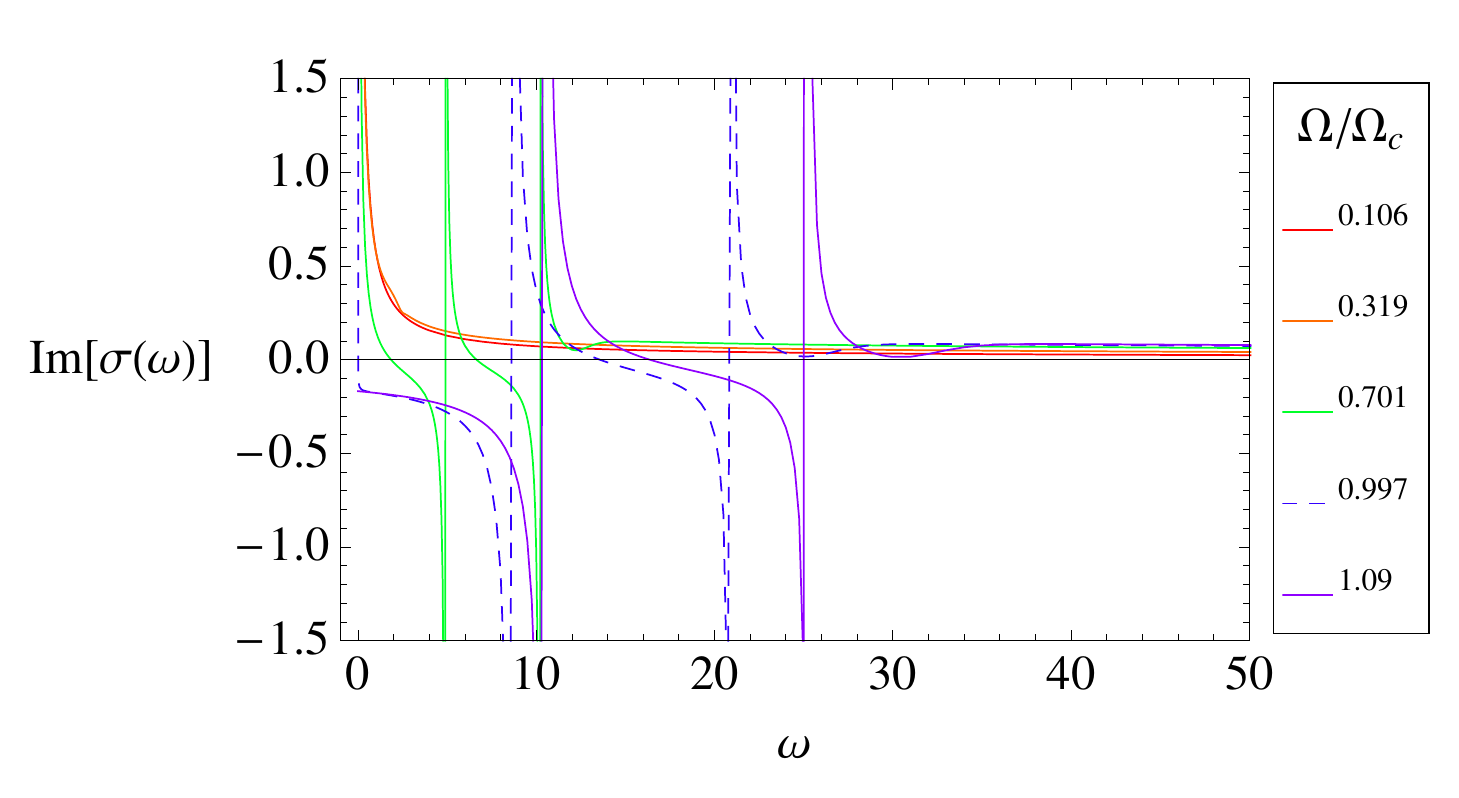, height=1.4 in} }
\caption{\em (a) A $2^{nd}$ order superfluid-insulator quantum phase transition at $\Omega_c=0.163$. (b) For $\Omega>\Omega_{c}$, we find an insulating gap. For $\half\Omega_{c}<\Omega<\Omega_{c}$, we find a superconducting pole, plus two isolated poles. For $\Omega<\half\Omega_{c}$, these poles merge with the zero-frequency superconducting pole.}
\label{soliton tune Om}
\end{figure}

In the black hole case, tuning $\Omega$ drove us through a superfluid-conductor phase transition at zero temperature.  The (zero-temperature) soliton shows the same effect, with the spontaneous condensate disappearing in a $2^{nd}$-order transition as $\Omega$ passes through a critical value, $\Omega_{c}\sim0.163$, as shown in Figure \ref{soliton tune Om}a.  Here, however, the normal phase is an insulator (cf Figure \ref{soliton tune Om}b) with the gap controlled by $(\Omega-\Omega_{c})$ and a double-pole structure as seen above.  Interestingly, this double-pole structure persists into the superconducting phase, with the gapped poles merging with the zero-frequency poles at a finite value of $\Omega\sim\half\Omega_{c}$. 

%

\subsection{Varying $m_G$ and the Gap}
\begin{figure}[!h]
\centerline{ \epsfig{figure=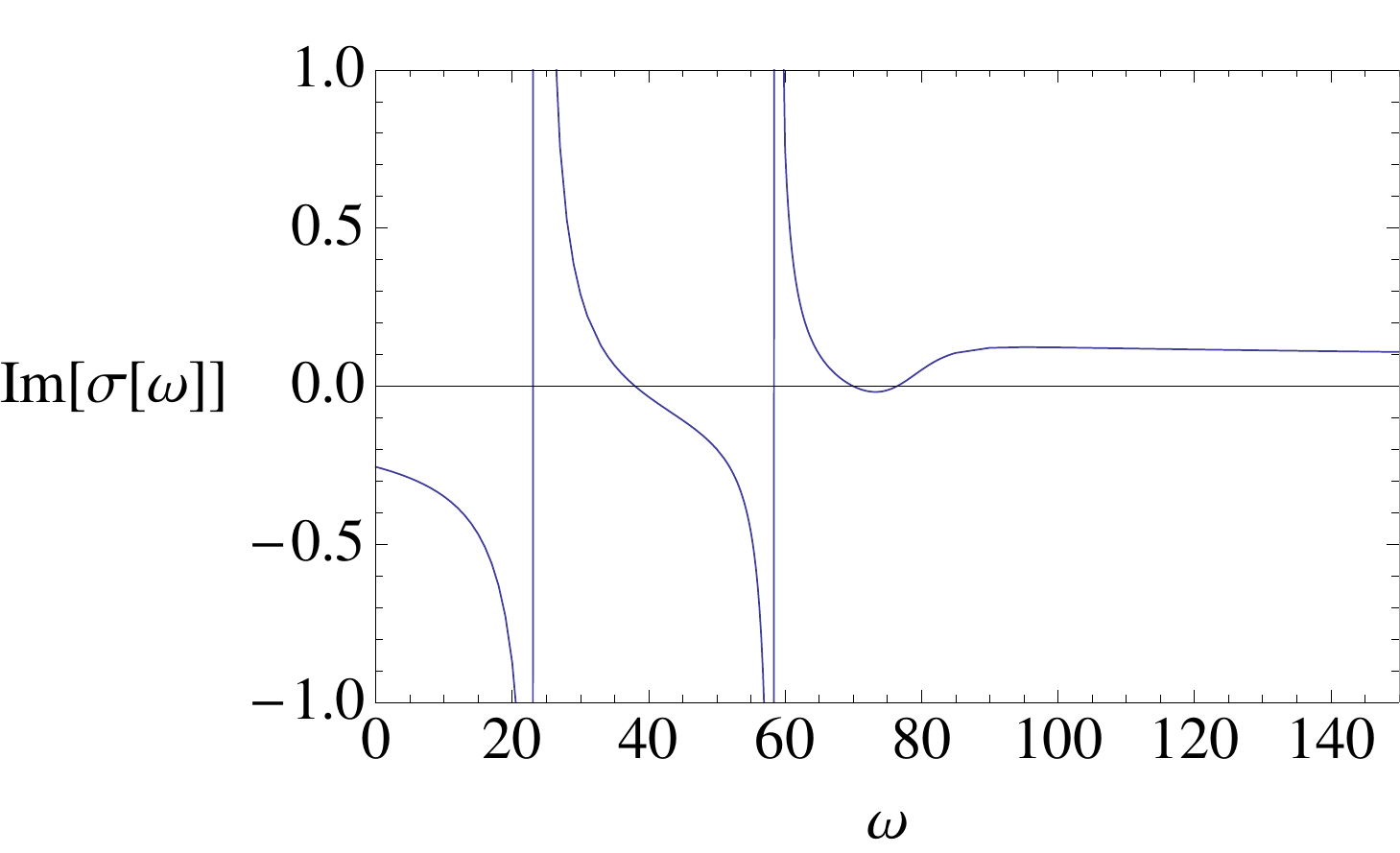, height=1.4 in}   \epsfig{figure=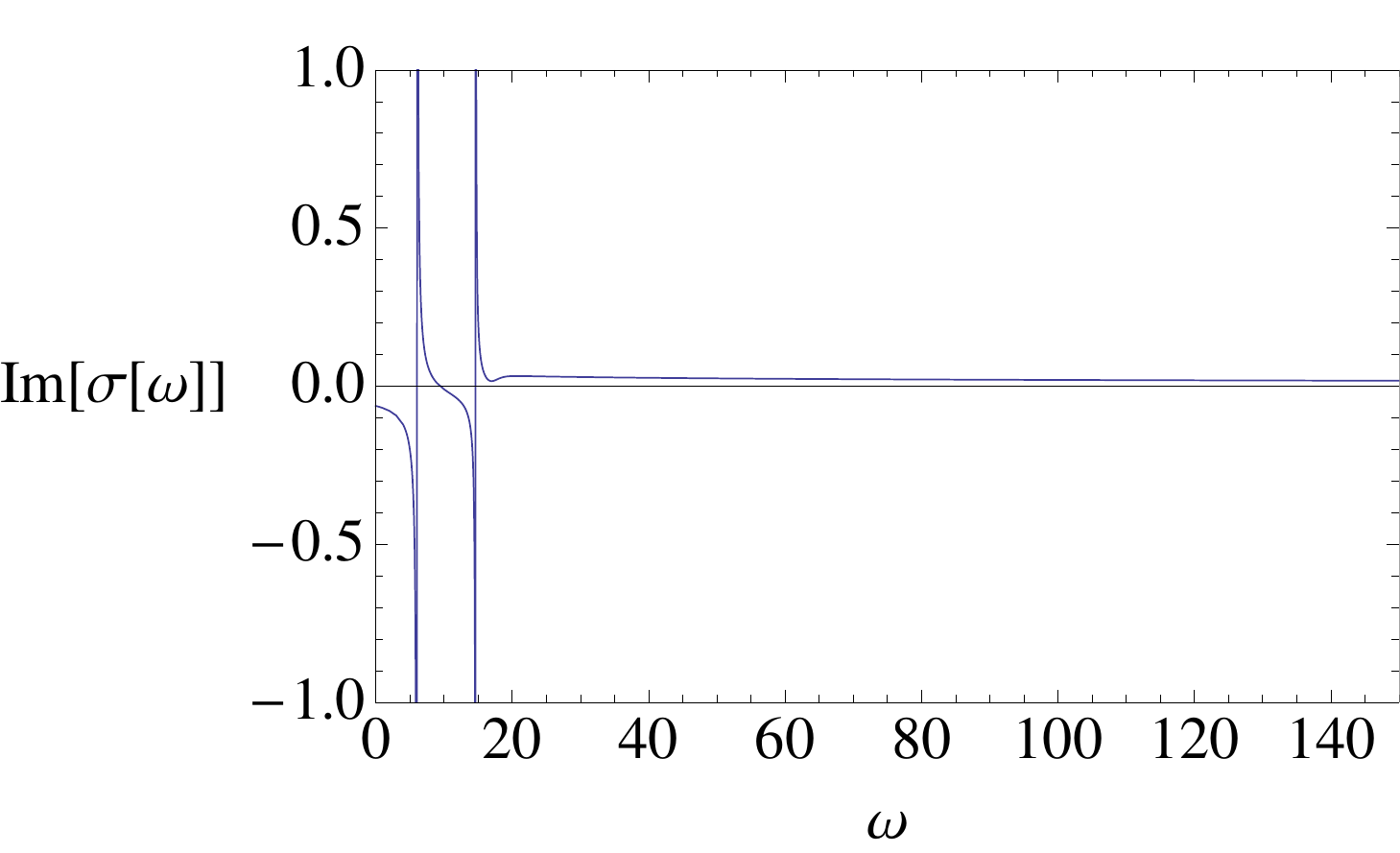, height=1.4 in} }
\caption{\em Varying $m_G$ rescales the insulating gap.  Here, $m_{G}$=$1\over4\pi$(left), $1\over16\pi$(right).}
\label{tune mG}
\end{figure}
In the above we have held $m_{G}$ fixed.  As it turns out, the only effect of varying $m_{G}$ is to rescale the gaps in all of the above (see Figure \ref{tune mG}).  This fits nicely with naive intuition for the effect of the compactification radius of the $\xi$-direction.

\section{Scaling Symmetries}
The system described by (\ref{EQ:ProbeAction}) and (\ref{EQ:pappalardoBH}) enjoys three distinct scaling symmetries which we can use to fix various parameters to convenient values.   Re-introducing $G_{N}$ and $R_{A}$ in the action and metric, the three scaling symmetries act as:
\begin{equation}
\begin{tabular}{c||c|c|c|c||c|c|c||c|c|c|c||c|c|c|}
  scaling symmetry & $t$ & $\xi$ & $x_i$ & $r$ & $T$ & $\Omega$ & $r_H$ & $ds^2$ & $\phi$ & $A_t$ & $A_\xi$ & $G_N$ & $q$ & $m$ \\ \hline
  $\alpha_{1}$ & $0$ & $0$ & $0$ & $0$ & $0$ & $0$ & $0$  & $-4$ & $0$ & $-2$ & $-2$ & $0$ & $2$ & $2$ \\
  $\alpha_{2}$ & $0$ & $0$ & $0$ & $0$ & $0$ & $0$ & $0$ & $0$ & $-2$ & $-2$ & $-2$ & $4$ & $2$ & $0$ \\
  $\alpha_{3}$ & $-2$ & $0$ & $-1$ & $-1$ & $2$ & $1$ & $-1$  & $0$ & $0$ & $2$ & $0$ & $0$ & $0$ & $0$ \\
 \end{tabular}\label{ScalingSymmetries1}
 \end{equation}
The first two symmetries can be used to fix $\frac{1}{16 \pi G_N}=1$ and $R_A=1$.  The third, which is the basic scaling symmetry of the Schr\"odinger system with dynamical exponent $z=2$, can be used to fix $r_{H}$ to a convenient reference value, $r_{o}$.  Given that $T=\frac{1}{\pi \Omega r_H^3}$, this fixes a relation between $T$ and $\Omega$.  To access more general values of these parameters, corresponding to $(r_H',T',\Omega')$,  we simply map the system to $(r_0,T,\Omega)=(\lambda^{-1} r_H', \lambda^{2} T', \lambda^{1} \Omega')$, with all physical parameters correspondingly rescaled.
(As usual, taking $q\to\infty$ while holding $q\Phi$ fixed for each matter field $\Phi$ gives us a probe limit in which backreaction is negligible \cite{Hartnoll:2008kx}.)

\end{document}